\documentclass[aps,prd,onecolumn,nofootinbib,12pt]{revtex4-2}

\usepackage[T1]{fontenc}
\usepackage[caption=false]{subfig}
\usepackage{lmodern,amsmath,amssymb,amsthm,mathtools,graphicx,makecell,float,xcolor,setspace,hyperref,bookmark}

\definecolor{dark-blue}{rgb}{0.3,0.3,0.7}
\definecolor{dark-red}{rgb}{0.7,0,0}
\hypersetup{
    colorlinks=true,
    linkcolor=dark-blue,
    citecolor=dark-red,
    urlcolor=dark-blue,
    linktoc=page
}

\newcommand{\+}{{\mkern2mu}}
\newcommand*{\di}{\mathrm{d}}
\newcommand*{\dd}{\mathrm{d}}

\newcommand{\abs}[1]{\mathopen{|}#1\mathclose{|}}
\renewcommand*{\Re}{\operatorname{Re}}
\renewcommand*{\Im}{\operatorname{Im}}
\newcommand*{\RP}{\mathbb{R}\mathrm{P}}
\newcommand*{\CP}{\mathbb{C}\mathrm{P}}
\newcommand{\Mpl}{M_\mathsf{Pl}}

\makeatletter
\newcommand{\lambdabarr}{{\mathchoice
  {\smash@bar\textfont\displaystyle{0.25}{1.2}\lambda}
  {\smash@bar\textfont\textstyle{0.25}{1.2}\lambda}
  {\smash@bar\scriptfont\scriptstyle{0.25}{1.2}\lambda}
  {\smash@bar\scriptscriptfont\scriptscriptstyle{0.25}{1.2}\lambda}
}}
\newcommand{\smash@bar}[4]{%
  \smash{\rlap{\raisebox{-#3\fontdimen5#10}{$\m@th#2\mkern#4mu\mathchar'26$}}}%
}
\makeatother

\begin{document}
\count\footins = 1000  
\setlength{\abovedisplayshortskip}{5pt plus 2pt minus 2pt}

\title{The Kontsevich-Segal Criterion in the No-Boundary State Constrains Anisotropy}

\author{Thomas Hertog$^\spadesuit$, Oliver Janssen$^\clubsuit$ and Joel Karlsson$^\spadesuit$ \vspace{0.1cm}}

\affiliation{$^\spadesuit${\small\slshape Institute for Theoretical Physics, KU Leuven, 3001 Leuven, Belgium} \\
  $^\clubsuit${\small\slshape Laboratory for Theoretical Fundamental Physics, EPFL, 1015 Lausanne, Switzerland}}

\begin{abstract} \vspace{0.15cm} \noindent
  We show that the Kontsevich-Segal-Witten (KSW) criterion applied to the no-boundary state constrains anisotropic deformations of de Sitter space.
  We consider squashed $S^3$ and $S^1 \times S^2$ boundaries and find that in both models, the KSW criterion excludes a significant range of homogeneous but anisotropic configurations.
  For squashed $S^3$ boundaries, the excluded range includes all surface geometries with negative scalar curvature, in line with dS/CFT reasoning.
  For $S^1 \times S^2$ boundaries, we find that KSW selects the low-temperature regime of configuration space where the $S^1$ is sufficiently large compared to the $S^2$.
  In both models, the KSW criterion renders the semiclassical wave function normalizable, up to one-loop effects.
\end{abstract}

\maketitle
\vspace{-0.4cm}
\tableofcontents
\pagenumbering{roman}
\clearpage
\newpage
\pagenumbering{arabic}

\section{Introduction} \label{introsec} \noindent
Kontsevich and Segal (KS) have argued that physically meaningful quantum field theories (QFTs) should be limits of theories that are well-defined on a class of ``allowable'' complex background geometries \cite{Kontsevich:2021dmb}.
The motivation for this criterion stems from the observation that complexifying a QFT in this way can account for the usual causality axiom that spacelike-separated operators commute in correlation functions.

Inspired by the path integral for free $p$-forms, KS defined the domain of allowable complex metrics as those satisfying
\begin{equation}
  \label{eq:KS_p-form_formulation}
  \Re(\sqrt{g}\+ g^{\mu_1 \nu_1} \hdots g^{\mu_p \nu_p} F_{\mu_1 \hdots \mu_p} F_{\nu_1 \hdots \nu_p}) > 0 \+ ,
\end{equation}
for all real, closed $p$-forms $F_{(p)} = \mathrm{d}A_{(p-1)}$ and all $p \in \{0, \hdots, D\}$, with $D$ the dimension of spacetime.
They also showed that any such metric can be diagonalized, locally, using a real change of basis, i.e.~that the metric can locally be written as
\begin{equation}
  \dd s^2 = g_{\mu\nu}\+ \dd x^\mu \dd x^\nu = \sum_{i=1}^D \lambda_i\+ (e^i)^2 \+ ,
\end{equation}
with $e^i = e_\mu{}^i \dd x^\mu$ real vielbeins and $\lambda_i \in \mathbb{C}$.
The collection of allowability criteria \eqref{eq:KS_p-form_formulation} for all $p$ then translate to
\begin{equation}
  \label{eq:KS_lambda_formulation}
  \sum_{i = 1}^D \+ \abs{\arg \lambda_i} < \pi\+,
\end{equation}
where $\arg(\cdot) \in (-\pi, \pi]$.
This formulation of the KS criterion shows that Lorentzian metrics lie on the boundary of the domain of allowable complex metrics.
Hence, a Lorentzian metric can be approached from within the domain of allowable metrics.
The idea is that \eqref{eq:KS_p-form_formulation} ensures absolute convergence of the matter path integral in this limit, thus specifying a sensible QFT.

Witten suggested that the KS criterion might play a role in quantum gravity and conjectured that physically meaningful saddle points of the gravitational path integral must be allowable backgrounds in the sense of KS \cite{Witten:2021nzp}.
Some evidence in favor of this conjecture comes from the fact that it eliminates certain pathological saddles, such as wormholes in flat space with vanishing action.
By contrast, certain complex solutions that appear physically sensible, such as the double cone of \cite{Saad:2018bqo}, pass the KSW criterion.

It is especially interesting to investigate the implications of KSW-allowability in combination with the no-boundary proposal for the wave function of the universe \cite{Hartle:1983ai}, in which complex metrics on compact manifolds feature abundantly (see e.g.~\cite{Jonas:2022uqb,Hertog:2023vot,Lehners:2023pcn,Maldacena:2024uhs} for recent work).
In this context KSW may well have real predictive power.
For instance, \cite{Hertog:2023vot} showed numerically that, under certain assumptions, KSW applied to the no-boundary saddles describing the origin of inflation selects those inflationary potentials for which the tensor-to-scalar ratio $r$ of primordial fluctuations is relatively small.
Subsequently, \cite{Maldacena:2024uhs,Janssen:2024vjn} provided an analytic explanation of this prediction.

Here we analyze to what extent KSW allows anisotropic deformations of de Sitter space, again in the framework of the no-boundary proposal.
To this end, we consider closed universes in four dimensions in pure gravity with a positive cosmological constant, with spatial slices $\Sigma_r$ that are either squashed three-spheres or products $S^1 \times S^2$.
Defining a Euclidean time coordinate $r$, the saddle-point metric can be written as
\begin{equation}
  \dd s_{\mathcal{M}}^2 = \di r^2 + \di s_{\Sigma_r}^2 \+.
\end{equation}
To see that KSW could significantly constrain the no-boundary wave function in this context, consider the action
\begin{equation} \label{theaction}
  S = - \underset{\mathcal{M}}{\int} \dd^4 x \sqrt{g} \left( \frac{\Mpl^2}{2} R - \Lambda \right) - \Mpl^2 \underset{\partial \mathcal{M}}{\int} \dd^3 y \sqrt{h}\+ K \+,
\end{equation}
where $\Mpl^2 = 1/(8\pi G_N)$.
Evaluating this on a saddle-point solution, using $R = 4 \Lambda/\Mpl^2$ and $K = \partial_r \log \sqrt{h}$, the on-shell action becomes
\begin{equation} \label{onshellaction}
  S = - \Lambda \underset{\mathcal{M}}{\int} \di^4 x \sqrt{g} - \Mpl^2 \, \partial_r \underset{\partial \mathcal{M}}{\int} \dd^3 y \sqrt{h} \+.
\end{equation}
The $0$-form condition in \eqref{eq:KS_p-form_formulation} amounts to the requirement that $\Re(\sqrt{g}) > 0$ everywhere on $\mathcal{M}$.
The KSW criterion therefore implies that the real part of the bulk term in \eqref{onshellaction} is negative for allowable saddles.
Since the real part of the boundary term typically provides a subleading contribution to the total action, at least for saddles corresponding to asymptotically classical spacetimes, this means we can expect KSW-allowable saddle points of the no-boundary wave function to be exponentially enhanced in the large volume limit.
Indeed, the real part of the boundary term in the action vanishes for asymptotically locally real de Sitter saddles of infinite volume in pure gravity since these are of the form $\dd s^2 = -\dd t^2 + e^{2t} ( h^{(0)}_{ij} + e^{-2t} h^{(2)}_{ij} + e^{-3t} h^{(3)}_{ij} + \hdots ) \dd x^i \dd x^j$ with $h^{(0)}_{ij}$ real and $h^{(2)}_{ij}$ determined by $h^{(0)}_{ij}$ (hence, real) from the Einstein equation \cite{deHaro:2000vlm}.%
\footnote{The argument given here can also be explicitly verified for no-boundary saddles corresponding to slow-roll inflationary histories.
  These saddles consist of an approximately Euclidean region glued onto an asymptotically Lorentzian, inflationary solution.
  The Euclidean region gives $\Re S = -12\pi^2 \Mpl^4/V_*$, up to slow-roll corrections, where $V_*$ is the potential energy at the south pole of the saddle, while $\int \dd^3 y \sqrt{h} \+ \Re K \sim a^3 \+ \Im\left( \dot{a}/a \right) \sim a^3 H_* \varepsilon_*/(aH_*)^3$, where $\Mpl^2 H_*^2 \sim V_*$, so $\Mpl^2 \int \dd^3 y \sqrt{h} \+ \Re K \sim \varepsilon_* \Mpl^4/V_*$ \cite{Janssen:2020pii,Maldacena:2024uhs,Janssen:2024vjn}.
  Hence, the contribution from the boundary term is slow-roll suppressed.}

Interestingly, in the squashed $S^3$ case, this would exclude anisotropic boundary configurations with negative three-curvature, since the corresponding saddles have Euclidean actions with positive real parts.
Both single and double squashings of the three-sphere can give rise to boundary configurations with negative Ricci scalar.
For a single squashing, there is one critical value of the squashing parameter at which the Ricci scalar changes sign.
For double squashings, the Ricci scalar turns out to be negative in a large patch of the configuration space.
This argument suggests, therefore, that KSW excludes those configurations, and this expectation is borne out by our detailed calculations below.

The exclusion of boundary configurations with negative curvature by KSW resonates with dS/CFT.
At least one version of the dS/CFT correspondence conjectures that the Hartle-Hawking wave function of the universe in the large volume regime can be specified in terms of the partition functions of dual Euclidean QFTs living on the surface where one evaluates the wave function \cite{Maldacena:2002vr,Hertog:2011ky,Anninos:2012ft,Hartle:2012qb}.
In this correspondence, the argument of the wave function relates to external sources in the dual partition function that turn on deformations.
The dependence of the partition function on the values of these sources, which include the three-geometry of the surface, yields a holographic no-boundary measure on the space of asymptotically locally de Sitter universes.
Furthermore, it turns out that Euclidean AdS/CFT generalized to complex deformations provides explicit (albeit non-realistic) examples of dS/CFT pairs in the semiclassical approximation \cite{Hertog:2011ky}.%
\footnote{This follows from the observation that all no-boundary saddles in low energy gravity theories with a positive scalar potential $V$ admit a geometric representation in which their weighting is fully specified by a complex AdS domain wall interior governed by an effective negative scalar potential.
  Quantum cosmology thus lends support to the view that Euclidean AdS/CFT and dS/CFT are two real domains of a single complexified theory.}

In this dS/CFT setup, the Ricci scalar on the conformal boundary enters as a ``mass'' term of conformally coupled scalars in the dual theory.
For geometries that are close to the round sphere, this is positive.
However, in regions of superspace corresponding to negative curvature configurations, the free dual theory is unstable.
This likely causes its partition function to diverge \cite{Witten1999,Bobev:2016sap}, thereby setting the amplitude of those configurations to zero.
Indeed one might say that the holographic form of the wave function indicates that regions of configuration space with negative Yamabe invariant%
\footnote{The Yamabe invariant $Y(\tilde h)$ is a property of conformal classes.
  It is essentially the infimum of the total scalar curvature in the conformal class of $\tilde h$, normalized with respect to the overall volume.
  There always exists a conformal transformation to a metric with constant scalar curvature, and the infimum defining $Y$ is obtained for this metric.
  Thus, the Yamabe invariant is negative in conformal classes containing a metric of constant $R < 0$.}
should be altogether excised from the minisuperspace%
\footnote{We stress that by ``minisuperspace'' in this paper we are referring to the restricted space of 3D boundary configurations, squashed three-spheres or products $S^1 \times S^2$, and the assumption that highly symmetric 4D saddles dominate $\Psi_\textsf{HH}$ in the semiclassical limit (cf.~footnote~\ref{MSPfootnote} and \cite{Halliwell:2018ejl}).
  We do not truncate the 4D gravity and integrate only over a select set of modes, with the exception of the limit of large $S^1/S^2$ ratio in \S\ref{S1S2sec} where this is effectively justified.}, in line with KSW.

It is intriguing that conformal classes with negative Yamabe invariant may well include the highly irregular constant density surfaces featuring in conventional semiclassical analyses of eternal inflation.
In those models of the dynamics of eternal inflation, one typically envisages cosmologies with a mosaic structure of bubble-like regions with negative curvature separated by inflationary domains.
The qualitative argument given above, based either on holography or on the KSW criterion, suggests that the amplitude of such highly irregular surfaces will be low or even zero in a proper quantum treatment of eternal inflation \cite{Hawking:2017wrd}.

\section{Squashed \texorpdfstring{$S^3$}{S3} boundary} \label{S3sec} \noindent
The first minisuperspace we consider is known as the Bianchi IX model and has a squashed $S^3$ boundary.
It was previously studied in e.g.~\cite{Jensen:1990xc,Bobev:2016sap,DiazDorronsoro:2018wro,Janssen:2019sex,Lehners:2024kus}.

Bianchi IX minisuperspace is a truncation of gravity with a positive cosmological constant to three degrees of freedom parametrizing the geometry of spatial slices that are double-squashed three-spheres, breaking $\mathfrak{so}(4) = \mathfrak{su}(2)_L \oplus \mathfrak{su}(2)_R$ to $\mathfrak{su}(2)_L$.
Instead of the single scale factor that specifies the size of the round $S^3$ in the FLRW solution, there are thus three parameters, viz.~the overall size and two squashing parameters.
The classical histories contained in this model are anisotropic Bianchi IX cosmologies that, with an appropriate choice of (dimensionless) time coordinate, can be written as
\begin{equation} \label{ds2double}
  H^2\+ \di s^2 = - \di t^2 + \frac{a(t)^2}{4} \biggl[ (\sigma^1)^2 + \frac{1}{1+\beta(t)} (\sigma^2)^2 + \frac{1}{1+\alpha(t)} (\sigma^3)^2  \biggr] \+,
\end{equation}
where $H^2 \equiv \Lambda/3 \Mpl^2$ and $\sigma^i$ are the left-invariant 1-forms on $S^3$ satisfying $2\+ \dd \sigma^i = - \varepsilon^i{}_{jk} \sigma^j \sigma^k$.
Explicitly, in terms of Euler angles ($0 \leq \theta \leq \pi$, $0 \leq \phi < 2\pi$, $\psi \cong \psi + 4\pi$), these can be written as
\begin{equation}
  \sigma^1 = \cos \psi\+ \dd \theta + \sin \psi \sin \theta\+ \dd \phi \+, \quad
  \sigma^2 = -\sin \psi\+ \dd \theta + \cos \psi \sin \theta\+ \dd \phi \+, \quad
  \sigma^3 = \dd \psi + \cos \theta\+ \dd \phi \+.
\end{equation}
Since the $\sigma^i$ are $\mathrm{SU}(2)_L$-invariant, the spatial slices are homogeneous ($\mathrm{SU}(2)_L$ acts transitively).
They are only isotropic for $\alpha = \beta = 0$, however, corresponding to a round $S^3$ of radius $a$.

One can truncate the theory further by setting either squashing parameter identically to zero, say $\beta \equiv 0$.
This is known as single-squashed or biaxial Bianchi IX minisuperspace.
In this case the $S^3$ is constructed as an $S^1$-bundle over $S^2$, with $\sqrt{1 + \alpha}$ specifying the relative size of the $S^2$ to the $S^1$, since $(\sigma^1)^2 + (\sigma^2)^2 = \dd \theta^2 + \sin^2 \theta\+ \dd \phi^2$ is the round metric on the unit $S^2$.
For convenience of finding analytic solutions, in the biaxial case we will write the metric as
\begin{equation}
  \label{eq:bb9_saddle}
  H^2\+ \dd s^2 = \frac{\dd r^2}{q(r)} + \frac{p(r)}{4} \bigl[(\sigma^1)^2 + (\sigma^2)^2\bigr] + \frac{q(r)}{4} (\sigma^3)^2 \+.
\end{equation}
In \S\ref{sec:bb9_NUT_KS}--\ref{sec:bb9_Bolt_KS} and the rest of this introduction we will study the biaxial case, returning to the double-squashed case in \S\ref{doublesquashedsec}.

The no-boundary wave function $\Psi_\textsf{HH}$ in the biaxial model is a function of the boundary values $P > 0$ and $Q > 0$ of both scale factors.
The semiclassical approximation of $\Psi_\textsf{HH} [P,Q]$ is specified by the on-shell action of no-boundary saddles of the form \eqref{eq:bb9_saddle} with $p(v) = P$ and $q(v) = Q$ at the boundary, which is located at a (complex) coordinate value $r = v$.%
\footnote{Strictly speaking, the validity and meaningfulness of the minisuperspace approximation we employ rests on the assumption that, in the full theory, the dominant saddle contributing to $\Psi_\textsf{HH} [P,Q]$ is of the most symmetric form \eqref{eq:bb9_saddle}, see also footnote~\ref{footnote:bb9_enhanced_saddle}.\label{MSPfootnote}}
With an ansatz of this form, the action \eqref{theaction} reduces to%
\footnote{We take $\sqrt{p^2} = p$ in the volume form.
  This follows from KSW, which implies one must select the saddle with $\Re(\sqrt{g}) > 0$.
  The choice of sign $\sqrt{p^2} = +p$ achieves this at the no-boundary south pole as one can check from what follows.}
\begin{equation}
  \label{eq:bb9_action}
  \begin{aligned}[b]
    S & = - \frac{2 \pi^2 \Mpl^2}{H^2} \biggl[ \int \dd r \biggl(\frac{q}{4p} (p')^2 + \frac{1}{2} p'q' - 3 p - \frac{q}{p} + 4\biggr)
    + \Bigl(q p' + \frac{1}{2} p q' \Bigr)\big|_{r=0} \biggr]
  \end{aligned}
\end{equation}
after integrating by parts.
The Hamiltonian constraint and the equation of motion obtained from varying $q$ read
\begin{equation}
  \label{eq:bb9_eoms}
  \frac{q}{4} (p')^2 + \frac{p}{2} p' q' + q + 3p^2 - 4p = 0 \+, \qquad
  p\+ p'' - \frac{1}{2} (p')^2 + 2 = 0 \+,
\end{equation}
while the equation of motion obtained from varying $p$ is redundant.
It is straightforward to verify that any solution to \eqref{eq:bb9_eoms} solves the full vacuum Einstein equation.
Without imposing boundary conditions yet, the most general solution to \eqref{eq:bb9_eoms} is
\begin{equation}
  \label{eq:bb9_general_solution}
  p(r) = p_0 + p_1 r + p_2 r^2 \+,\qquad
  q(r) = \frac{c_0 + c_1 r + (4 - 6 p_0)r^2 - 2 p_1 r^3 - p_2 r^4}{p(r)} \+,
\end{equation}
where the coefficients are subject to the constraints
\begin{equation}
  \label{eq:bb9_coeff_constraints}
  p_1^2 = 4 (1 + p_0 p_2) \+,\qquad
  2 p_0 (4 - 3 p_0) + 2 c_0 p_2 - c_1 p_1 = 0 \+.
\end{equation}
We are left with three undetermined coefficients at this stage.

We are interested in compact, regular saddles with a single boundary.
Placing the south pole at the origin $r=0$, we thus look for regular solutions with $p(0) = 0$ or $q(0) = 0$.%
\footnote{One can also obtain a no-boundary saddle on the cross-cap $\RP^4 \setminus B^4$ without setting $p(0) = 0$ or $q(0) = 0$ and instead taking a $\mathbb{Z}_2$-quotient \cite{Daughton:1998aa}.
  We will not consider this possibility here.}
Taking $p(0) = 0$, regularity at $r = 0$ requires $c_0 = c_1 = 0$ to avoid a curvature singularity, as seen from $R_{\mu\nu\rho\sigma} R^{\mu\nu\rho\sigma}$.
This solves the second constraint in \eqref{eq:bb9_coeff_constraints}, and the first one is solved by $p_1 = \pm 2$.
By using the $r \to -r$ symmetry, we may take $p_1 = 2$ without loss of generality.
Thus, we have arrived at the one-parameter family of solutions
\begin{equation}
  \label{eq:bb9_NUT}
  p(r) = 2 r + p_2 r^2 \+,\qquad
  q(r) = \frac{4 r - 4 r^2 - p_2 r^3}{2 + p_2 r} \+,
\end{equation}
known as the Taub-NUT-dS solutions.
Expanding this solution around $r = 0$, we see that it behaves as a standard no-boundary saddle \cite{Hartle:1983ai}, i.e.~the three-spheres become round and the topology of the interior solution is that of the four-ball $B^4$.

Requiring instead $q(0) = 0$ but $p(0) = p_0 \neq 0$, we need $c_1 \neq 0$ to obtain a compact solution.
Near the south pole, the metric then becomes
\begin{equation}
  H^2\+ \dd s^2 = \biggl(\frac{p_0}{c_1} \frac{\dd r^2}{r} + \frac{c_1 r}{4 p_0} (\sigma^3)^2 + \frac{p_0}{4} \bigl[(\sigma^1)^2 + (\sigma^2)^2\bigr]\biggr) \bigl[1 + \mathcal{O}(r)\bigr] \+.
\end{equation}
Changing coordinates to $\dd \tau^2 = p_0 \dd r^2 / (c_1 r)$ we see that this is, locally, $\mathbb{R}^2$ fibered over a round, finite-radius $S^2$.
To avoid a conical singularity we need
\begin{equation}
  c_1 = 2 p_0 \+,
\end{equation}
where we have again used the $r \to - r$ symmetry.
The constraints \eqref{eq:bb9_coeff_constraints} can now be used to express $p_{1,2}$ in terms of $p_0$.
This yields another one-parameter family of solutions given by \eqref{eq:bb9_general_solution} with
\begin{equation}
  \label{eq:bb9_Bolt}
  \begin{aligned}
     & c_0 = 0 \+,               && c_1 = 2 p_0 \+,                   \\
     & p_1 = 4 - 3 p_0 \+,\qquad && p_2 = \frac{p_1^2 - 4}{4 p_0} \+,
  \end{aligned}
\end{equation}
known as the Taub-Bolt-dS solutions, with $\CP^2 \setminus B^4$ topology.

\subsection{KSW constraints on single-squashed Taub-NUT-dS} \label{sec:bb9_NUT_KS} \noindent
So far, we have only considered the no-boundary conditions at $r = 0$ but have not yet imposed the final boundary conditions $p(v) = P$ and $q(v) = Q$.
As we will see, imposing final boundary conditions generically leads to complex solutions \cite{Jensen:1990xc,DiazDorronsoro:2018wro,Janssen:2019sex,Lehners:2024kus,Daughton:1998aa}, raising the question whether they satisfy the KSW criterion \eqref{eq:KS_lambda_formulation}.
We now analyze KSW-allowability in the large volume limit, first for the Taub-NUT-dS saddles and next for the Bolts.

Since the squashing parameter
\begin{equation}
  \label{eq:bb9_alpha_def}
  \alpha \equiv \frac{P}{Q} - 1 > -1
\end{equation}
tends to a constant in asymptotically classical histories \cite{DiazDorronsoro:2018wro}, we implement the large volume limit by taking $P$ large while keeping $\alpha$ fixed.

Writing $r = \gamma(\ell)$ with $\gamma$ a curve in the complex $r$-plane and $\ell$ a real coordinate, the KSW constraint \eqref{eq:KS_lambda_formulation} on the metric \eqref{eq:bb9_saddle} reads
\begin{equation}
  \label{eq:bb9_KSW}
  \biggl|\arg \frac{\gamma'^{\+ 2}}{q(\gamma)}\biggr| + \abs{\arg q(\gamma)} + 2 \abs{\arg p(\gamma)} < \pi \+.
\end{equation}
Given a saddle $[p(r)$, $q(r)]$ and an endpoint $r = v$ where the final boundary conditions are met, \eqref{eq:bb9_KSW} becomes a pointwise constraint on $\gamma'$.
To determine allowability, we will consider a marginally allowable, extremal curve $\gamma_\mathsf{e}$ saturating the inequality \eqref{eq:bb9_KSW}.
Points where $p$ or $q$ vanish, such as the south pole, are special since at those points the second and third term in \eqref{eq:bb9_KSW} depend on $\gamma'$.
Away from those points, a vertical curve $\gamma' \in i \mathbb{R}$ is marginally allowable only where $\arg p = 0$ since the strict criterion becomes $\pi + 2 \abs{\arg p(\gamma)} < \pi$.
A strictly allowable curve is thus everywhere right-moving or everywhere left-moving; it cannot turn around.
Writing $\gamma(\ell) = \ell \+ e^{i \theta}$ close to the origin, where $p(r) = 2 r + \mathcal{O}(r^2) = q(r)$, the KSW criterion becomes $\abs{\theta} < \pi/4$.
Hence, only right-moving curves can be allowable and we impose this also on the extremal curves $\gamma_\mathsf{e}$ (that could otherwise turn around where $p \in \mathbb{R}^+$) in order to be able to get arbitrarily close to $\gamma_\mathsf{e}$ with strictly allowable curves.

Imposing the final boundary conditions $p(v) = P > 0$ and $q(v) = Q > 0$ on the Taub-NUT solution implies that
\begin{equation}
  \label{eq:bb9_NUT_p2_and_v}
  p_2 = \frac{P - 2v}{v^2} \+,\qquad
  2v^3 + (P-4)v^2 + PQ = 0 \+.
\end{equation}
At large volume and fixed squashing $\alpha$, this cubic equation for $v$ has one negative real solution $v = -P/2 + 2 \alpha / (1 + \alpha) + \mathcal{O}(P^{-1})$ and two complex conjugate solutions%
\footnote{When necessary, we will use subleading terms in the expression for $v$ as well.
  These are found by starting from this expression and solving \eqref{eq:bb9_NUT_p2_and_v} perturbatively in $1/P$.}
\begin{equation}
  \label{eq:bb9_NUT_v}
  v = \left( \frac{1}{1 + \alpha} \pm i \sqrt{\frac{P}{1 + \alpha}} \right) \bigl[1 + \mathcal{O}(P^{-1})\bigr] \quad \text{as } P \to \infty \+.
\end{equation}
The real part of the Euclidean on-shell action \eqref{eq:bb9_action} of the former solution is%
\footnote{The on-shell Lagrangian has a pole at $r = -2/p_2$ where $p(r) = 0$.
  For the real solution for $v$, this pole lies on the real axis between the origin and $v$.
  However, the residue of the on-shell Lagrangian is zero so the pole can be avoided on either side with the same result.}
$\Re S = - (\pi^2 \Mpl^2 / 2H^2)P^2 + \mathcal{O}(P)$ while for the complex conjugate saddles
\begin{equation}
  \label{eq:bb9_NUT_ReS}
  \Re S = - \frac{4 \pi^2 \Mpl^2}{H^2} \frac{1 + 2 \alpha}{(1 + \alpha)^2} + \mathcal{O}(P^{-1}) \+.
\end{equation}
Hence, the saddle with negative real $v$ has the highest semiclassical weight $\exp(-2\Re S)$.%
\footnote{\label{footnote:bb9_enhanced_saddle}%
  Curiously, for $\alpha = 0$, this third saddle is semiclassically enhanced over the standard Hartle-Hawking saddle even though it breaks the $O(4)$-symmetry in the interior of the solution.}
This saddle does not predict Lorentzian histories since its on-shell action is purely real, however \cite{PhysRevD.41.1815,Hartle:2008ng}.
Additionally, it does not satisfy the KSW criterion since, as explained above, any allowable curve $\gamma$ is right-moving and, thus, cannot end in the left half-plane.
This provides another example in which KSW excludes clearly unphysical saddles, adding to the list in \cite{Witten:2021nzp}.

The question now becomes whether the physically interesting saddles with complex endpoints \eqref{eq:bb9_NUT_v} are allowable by KSW in the large $P$ limit.
Since these saddles are related by complex conjugation, either both satisfy the KSW criterion or neither does.
Going forward, we focus on the one with $v$ in the first quadrant.
We will first exclude a range of $\alpha$ by an integrated version of the 0-form criterion and then consider the extremal curve $\gamma_\mathsf{e}$, which will exclude an even larger range.

Integrating the 0-form criterion in \eqref{eq:KS_p-form_formulation} over the saddle gives
\begin{equation}
  \label{eq:bb9_KS0_integrated}
  \int \dd^4 x \Re(\sqrt{g}) = 2 \pi^2 \int \dd r \+ p(r) > 0 \+,
\end{equation}
where we have used that \eqref{eq:KS_lambda_formulation} implies $\sqrt{p^2} = +p \sim 2r$ near $r = 0$.
For the Taub-NUT solution \eqref{eq:bb9_NUT}, using \eqref{eq:bb9_NUT_p2_and_v} and \eqref{eq:bb9_NUT_v}, this becomes
\begin{equation}
  0 < \Re\bigl(v(P + v)\bigr)
  = 2 \frac{1 + 2 \alpha}{(1 + \alpha)^2} + \mathcal{O}(P^{-1}) \+.
\end{equation}
We note that this holds precisely when $\Re S < 0$ (up to $1/P$ corrections), see \eqref{eq:bb9_NUT_ReS}, i.e.~for $\alpha > -1/2$.
This confirms the expectation that the Gibbons-Hawking-York (GHY) boundary term contributes subleadingly to $\Re S$, cf.~our comments about this in \S\ref{introsec} (one can also verify this explicitly: the real part of the GHY term vanishes like $1/P$ as $P \to \infty$).

Having excluded $\alpha < -1/2 + \mathcal{O}(P^{-1})$, we turn to the extremal curve to analyze $\alpha > -1/2$.
The endpoint $v$ has a finite $\mathcal{O}(P^0)$ real part, but a large $\mathcal{O}(P^{1/2})$ imaginary part.
For large $P$, $p(r)$ and $q(r)$ are approximately real and positive on the interval $r \in (0, \Re v)$ since $p_2$ is real up to $1/P$ corrections.
Thus, a curve $\gamma$ along the real axis is allowable until the vertical through $v$.
Instead, the way in which KSW might fail is if the up-moving extremal curve $\gamma_e$ passes below, rather than to the left, of the endpoint $v$.
Close to the origin, \eqref{eq:bb9_KSW} is saturated by the curve $\gamma_\mathsf{e}$ defined by $\gamma_\mathsf{e}' = i/p(\gamma_\mathsf{e})$ that starts at an angle $\arg \gamma_\mathsf{e}'(0) = \pi/4$.
This equation continues to hold as long as $\gamma_\mathsf{e}$ does not intersect a line where $p \in \mathbb{R}$ or $q \in \mathbb{R}$, in which case some absolute values in \eqref{eq:bb9_KSW} should be treated differently, or a line where $p \sqrt{q} \in i \mathbb{R}$, in which case there is no allowable direction $\gamma'$.
If such issues do not occur, the extremal curve can be obtained by integrating $\gamma_\mathsf{e}' = i/p(\gamma_\mathsf{e})$ and hence satisfies
\begin{equation}
  \label{eq:bb9_NUT_extremal_curve}
  \gamma_\mathsf{e}(\ell)^2 + \frac{P - 2 v}{3 v^2} \gamma_\mathsf{e}(\ell)^3 = i \ell \+,
\end{equation}
where we have used $\gamma_\mathsf{e}(0)=0$, \eqref{eq:bb9_NUT} and \eqref{eq:bb9_NUT_p2_and_v}.
We will sometimes refer to this curve as the ``naive'' extremal curve, since it is the one obtained by dropping the absolute value signs in \eqref{eq:bb9_KSW}.
Note that the naive extremal curve saturates the 0-form criterion by construction, whereas the assumptions $0 \leq \arg p(\gamma_\mathsf{e})$, $0 \leq \arg q(\gamma_\mathsf{e})$ and $\arg q(\gamma_\mathsf{e}) + 2 \arg p(\gamma_\mathsf{e}) \leq \pi$ guarantee that the higher-form criteria are satisfied.

Intersections of $\gamma_\mathsf{e}$ with lines where $p \in \mathbb{R}$, $q \in \mathbb{R}$ or $p \sqrt{q} \in i \mathbb{R}$ are studied in detail in Appendix~\ref{app:bb9_NUT_extremal_curve_validity} in the large $P$ limit.
Fig.~\ref{fig:bb9_NUT_examples} illustrates how $\gamma_\mathsf{e}$ differs between the $\alpha < -1/6$ and $\alpha > -1/6$ cases.
We find that $p \sqrt{q} \in i \mathbb{R}$ is never encountered and
\begin{itemize}
  \item $-1 < \alpha < -1/2$: $\gamma_\mathsf{e}$ intersects $p \in \mathbb{R}$ below and to the right of the endpoint $v$.
        Hence we conclude that these saddles are not allowable.%
        \footnote{\label{footnote:bb9_NUT}%
          This holds even if $\gamma_\mathsf{e}$ intersects $q \in \mathbb{R}$ since, after that point, the true extremal curve would be less steep than the naive one.}
        This is precisely the range of $\alpha$ that we excluded above by integrating the 0-form criterion, which is not surprising since $\gamma_\mathsf{e}$ saturates this criterion by construction.

  \item $-1/2 < \alpha < -1/6$: $\gamma_\mathsf{e}$ intersects $q \in \mathbb{R}$ at a distance $\mathcal{O}(P^0)$ from the origin, beyond which \eqref{eq:bb9_NUT_extremal_curve} is no longer valid.
        The 0-form criterion is, however, satisfied.
        In this regime we resort to numerics to find the actual extremal curve as opposed to the naive one.
        We find strong evidence that no $\alpha \in (-1/2, -1/6)$ is allowable in the large $P$ limit, see Fig.~\ref{fig:numerics}.

  \item $-1/6 < \alpha$: $\gamma_\mathsf{e}$ does not intersect $p \in \mathbb{R}$ or $q \in \mathbb{R}$ before passing to the left of $v$.
        Thus, these saddles are allowable.
\end{itemize}

\begin{figure}[ht!]
  \centering
  \includegraphics[width=340pt]{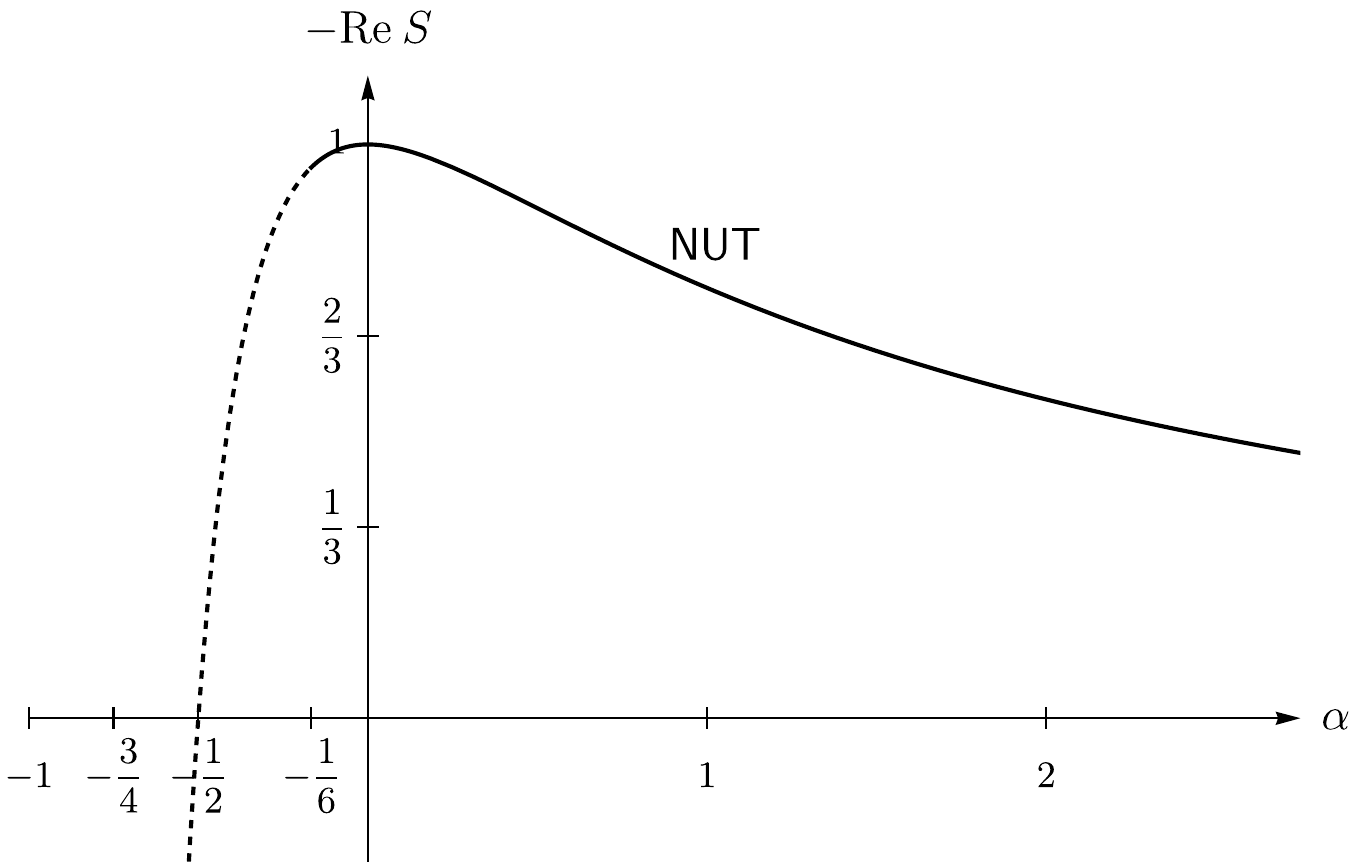}
  \caption{The KSW criterion applied to the no-boundary state excludes strongly anisotropic cosmologies on the Taub-NUT-dS branch in biaxial Bianchi IX minisuperspace.
    Shown is the real part of the Euclidean action $S$ (in units of $4\pi^2 \Mpl^2/H^2$) of NUT-type no-boundary saddles as a function of the squashing $\alpha$ of their $S^3$ boundary, in the large volume limit.
    The squashing parameter $\alpha$ is a measure of the size of the $S^2$ base relative to the $S^1$ fiber, with $\alpha =0$ corresponding to the round $S^3$.
    Only boundary configurations with $\alpha > -1/6$ are allowed by the KSW criterion (indicated by a solid line).
    Other notable values are $\alpha = -1/2$, where $\Re S$ changes sign, and $\alpha = -3/4$, where the curvature of the boundary $R^{(3)} \propto (3+4\alpha)/(1+\alpha)$ changes sign.}
  \label{fig:bb9_NUT_results}
\end{figure}

Notice that the Ricci curvature of the single-squashed boundary $S^3$ becomes negative when $\alpha < -3/4$.
One can calculate this from \eqref{eq:bb9_saddle}, or by using \eqref{R3boundary} below for the double-squashed boundaries, setting $\beta = 0$.
In summary, the KSW criterion applied to the no-boundary state constrains the Taub-NUT-dS branch in this minisuperspace from $\alpha \in (-1,\infty)$ to $\alpha \in (-1/6,\infty)$, excluding, among others, the entire range of boundary configurations with negative three-curvature ($\alpha < -3/4$) and positive real part of the action ($\alpha < -1/2$).
Fig.~\ref{fig:bb9_NUT_results} illustrates these results.

\subsection{KSW constraints on single-squashed Taub-Bolt-dS} \label{sec:bb9_Bolt_KS} \noindent
We now turn to the allowability of the Taub-Bolt-dS solutions, given by \eqref{eq:bb9_general_solution} and \eqref{eq:bb9_Bolt} in the large $P$ limit, with fixed squashing $\alpha$.
As above, we start by excluding some regions of $\alpha$ by considering e.g.~the integrated 0-form criterion and then turn to a more refined analysis employing the extremal curve.

The first step is to fix the parameter $p_0$ of the solution and the endpoint $v$ in terms of the final boundary data $P$ and $Q$.
Imposing the final boundary condition $p(v) = P$ gives
\begin{equation}
  \label{eq:bb9_Bolt_p0eq}
  P = p_0 + (4 - 3 p_0) v + \frac{(4 - 3 p_0)^2 - 4}{4 p_0} v^2 \+,
\end{equation}
which is solved by
\begin{equation}
  p_0 = \frac{2}{(2 - 3v)^2} \biggl(P - 2 v (2 - 3v) \pm \sqrt{P^2 - 4 v (2 - 3v) P + v^2 (2 - 3v)^2}\biggr) \+.
\end{equation}
Imposing the second boundary condition $q(v) = Q = P/(1 + \alpha)$ gives an equation for $v$ involving the above square root.
Solving for the square root and squaring, this becomes a single polynomial equation for $v$ independent of the $\pm$ sign above.
Naively, the polynomial has degree eight but there is a false root at $v=2/3$.
The actual roots are the solutions to the following seventh degree polynomial in $v$:
\begin{equation}
  \label{eq:bb9_Bolt_veq}
  \begin{aligned}[b]
    0 & = (3v - 2) P^4 + 2 (1+\alpha) v \bigl((3v-4)v + 2\bigr) P^3 + (1+\alpha)^2 v^3 \bigl(3(v-2)v + 4\bigr) P^2
    \\ & \quad
       - 16 (1+\alpha)^2 v^3 (v-1)^2 P - 4 (1+\alpha)^2 v^4 (3v-2) (v-1)^2 \+,
  \end{aligned}
\end{equation}
where we have collected powers of $P$ to facilitate finding the roots perturbatively in $1/P$.
There is one real root at $v = 2/3 - 8(1+\alpha)/(27 P) + \mathcal{O}(P^{-2})$, two real roots at $v = \pm P/2 + \mathcal{O}(P^0)$ and two pairs of complex conjugate roots, of which the ones in the upper half plane are
\begin{equation}
  \label{eq:bb9_bolt_pm}
  \begin{aligned}
     & v = i\sqrt{\frac{P}{1+\alpha}} + \frac{1}{3} \Biggl(1 \pm \frac{\sqrt{\alpha^2 - 10 \alpha - 2}}{1+\alpha}\Biggr) + \mathcal{O}(P^{-1/2}) \+, \\
     & p_0 = \frac{4}{3} - \frac{2}{9} (1+\alpha) \Biggl(1 \pm \frac{\sqrt{\alpha^2 - 10 \alpha - 2}}{1+\alpha}\Biggr) + \mathcal{O}(P^{-1}) \+,
  \end{aligned}
\end{equation}
where the signs in the two expressions are correlated.%
\footnote{Further subleading corrections to these expressions are obtained through \eqref{eq:bb9_Bolt_p0eq} and \eqref{eq:bb9_Bolt_veq} and will be used whenever necessary.}

Before turning to the cosmologically relevant complex conjugate saddles, we comment on the three real solutions.
The solution with $v = 2/3 + \mathcal{O}(P^{-1})$ is regular and real Euclidean if $\gamma$ is taken along the real axis.
It hence satisfies KSW.
However, even though its volume is positive, its on-shell action is real and positive, $S = 27 \pi^2 \Mpl^2 P^3 / [8H^2 (1+\alpha)^2] + \mathcal{O}(P^0)$, due to a large contribution from the boundary term.
It is thus semiclassically suppressed with vanishing weight in the large $P$ limit and we will not consider it further.

The solutions with $v = \pm P/2 + \mathcal{O}(P^0)$ have real on-shell actions $S = \mp 9 \pi^2 \Mpl^2 P^3 / (4 H^2) + \mathcal{O}(P^2)$, which means that one of them is semiclassically enhanced.
However, neither satisfies KSW.
This is immediate for the one with $\Re v < 0$ since, as for the NUTs, allowable curves must be everywhere right-moving.
For the solution with $v = P/2 + \mathcal{O}(P^0)$, one can show that the origin is separated from the endpoint $v$ by two lines on which $p(r) \in i \mathbb{R}$, passing through $r = 1 + \mathcal{O}(P^{-1})$ and $r = P/4 + \mathcal{O}(P^0)$, respectively, and extending to infinity.
In the region in between these lines, there are no allowable directions $\gamma'$ since there $2 \abs{\arg p(r)} > \pi$ and the allowable directions have to satisfy \eqref{eq:bb9_KSW}.

Going forward, we only consider the cosmologically relevant solutions in \eqref{eq:bb9_bolt_pm}, which we will refer to as the plus- and minus-Bolts (for the upper/lower sign).
Their complex conjugates are allowable if and only if they are.
Note that the square roots in \eqref{eq:bb9_bolt_pm} are imaginary when $5-3\sqrt{3} < \alpha < 5+3\sqrt{3}$ and real otherwise.

As explained above, allowable curves must be everywhere right-moving.
Considering $v$ in \eqref{eq:bb9_bolt_pm} we see that the endpoint of the plus-Bolt is always in the right half-plane whereas the endpoint of the minus-Bolt is only in the right half-plane for $\alpha > -1/4$.
Thus, the minus-Bolts with $\alpha < -1/4$ do not satisfy KSW in the large $P$ limit.

The analysis close to the origin is different from the NUT case since now $p(r) = p_0 + \mathcal{O}(r)$ with $p_0 \neq 0$.
Writing $\gamma(\ell) = \ell \+ e^{i\theta} + \mathcal{O}(\ell^2)$, the KSW criterion in the $\ell \to 0$ limit thus becomes
\begin{equation}
  \abs{\theta} < \frac{\pi}{2} - \abs{\arg p_0} \+.
\end{equation}
If $\abs{\arg p_0} \geq \pi/2$ we can immediately conclude that the saddle violates KSW, since there is no allowable direction $\gamma'$ to start from at $r = 0$.
This excludes $\alpha > 5$ for both Bolts in the large $P$ limit, which is seen from the sign of the real part of $p_0$ in \eqref{eq:bb9_bolt_pm}.

Before turning to the extremal curve, we also consider the integrated 0-form criterion \eqref{eq:bb9_KS0_integrated}.
When $5-3\sqrt{3} < \alpha < 5+3\sqrt{3}$, both Bolts satisfy this criterion.
Outside this region, the minus-Bolt satisfies it whereas the plus-Bolt does not if $\alpha$ is either too large or too close to $-1$.
The only bound that improves on what we found above is the lower bound on $\alpha$.
Concretely, this reads
\begin{equation}
  (1+\alpha)^2 (17-\alpha) - (\alpha^2 - 10\alpha -2)^{3/2} > 0 \+,
\end{equation}
where $\alpha < 5 - 3\sqrt{3}$ so that $\alpha^2 - 10\alpha - 2 > 0$.
Solving for $\alpha$ gives $\alpha > \alpha_1$ where
\begin{equation}
  \label{eq:bb9_+bolt_alpha_1}
  5 \alpha_1^4 - 68 \alpha_1^3 - 81 \alpha_1^2 - 46 \alpha_1 - 11 = 0 \+ ,
  \qquad \alpha_1 \approx -0.468 \+ .
\end{equation}
Since these saddles are approximately Lorentzian along the vertical through $v$, we expect this to exclude $\Re S > 0$ since the boundary term should contribute subleadingly to $\Re S$.
Computing the on-shell action, we see that this is indeed the case:
\begin{equation}
  \Re S = \frac{4 \pi^2 \Mpl^2}{27 H^2} \biggl(\alpha - 17 \pm \frac{\Re\bigl((\alpha^2 - 10 \alpha - 2)^{3/2}\bigr)}{(1+\alpha)^2}\biggr) \+.
\end{equation}

For the minus-Bolt, the region that has not been excluded so far is $-1/4 < \alpha < 5$.
In this region, we analyze the extremal curve $\gamma_\mathsf{e}$ to determine allowability, but this turns out not to impose further constraints.
The analysis differs depending on whether the square root in \eqref{eq:bb9_bolt_pm} is real or not, i.e.\ on whether $\alpha$ is greater or less than $5-3\sqrt{3}$.
When $\alpha > 5 - 3\sqrt{3}$, $\arg p_0 > 0$ and the extremal curve is given by $\gamma_\mathsf{e}' = i/p(\gamma)$ close to the origin, which gives
\begin{equation}
  \label{eq:bb9_bolt_gamma_1}
  p_0 \gamma_\mathsf{e}(\ell) + \frac{4 - 3p_0}{2} \gamma_\mathsf{e}(\ell)^2 + \frac{4 - 8 p_0 + 3 p_0^2}{4 p_0} \gamma_\mathsf{e}(\ell)^3 = i \ell \+.
\end{equation}
When $\alpha < 5 - 3\sqrt{3}$, on the other hand, $\arg p_0$ vanishes to leading order in large $P$ but a subleading correction gives $\arg p_0 < 0$.
In this case, the extremal curve is given by $\gamma_\mathsf{e}' = i p(\gamma)$, which gives
\begin{equation}
  \label{eq:bb9_bolt_gamma_2}
  \frac{1}{2} \log\biggl(\frac{6 p_0 + (18 - 9 p_0) \gamma_\mathsf{e}(\ell)}{6 p_0 + (6 - 9 p_0) \gamma_\mathsf{e}(\ell)}\biggr) = i \ell \+.
\end{equation}
However, $\gamma_\mathsf{e}$ crosses a line on which $p \in \mathbb{R}$ at a point with $\Re \gamma_\mathsf{e} = \mathcal{O}(P^{-3})$ and $\Im \gamma_\mathsf{e} = \mathcal{O}(P^{-3/2})$, beyond which $\gamma_\mathsf{e}' = i/p(\gamma_\mathsf{e})$ is the correct equation.
This results in an additional, constant term in \eqref{eq:bb9_bolt_gamma_1} which can be neglected when analyzing $\gamma_\mathsf{e}$ to the order we need in $1/P$.
Using the methods explained in detail in Appendix~\ref{app:bb9_NUT_extremal_curve_validity}, we find that
\begin{itemize}
  \item $-1 < \alpha < -1/4$: the minus-Bolt is not allowable since $\Re v < 0$.

  \item $-1/4 < \alpha < 5 - 3\sqrt{3}$: after exiting the region with $\arg p < 0$ close to the origin, $\gamma_\mathsf{e}$ remains in a region in which $\arg p > 0$, $\arg q > 0$ and $2 \arg p + \arg q < \pi$ until passing to the left of the endpoint $v$.
        Hence, the minus-Bolt is allowable.

  \item $5 - 3\sqrt{3} < \alpha < 5$: the extremal curve never crosses any line on which $p \in \mathbb{R}$, $q \in \mathbb{R}$ or $p \sqrt{q} \in i \mathbb{R}$ before passing to the left of the endpoint $v$ and we conclude that the minus-Bolt is allowable.

  \item $5 < \alpha$: the minus-Bolt is not allowable since there is no allowable direction $\gamma'$ at the origin.
\end{itemize}

For the plus-Bolt, the region that has not been excluded so far is $\alpha_1 < \alpha < 5$ with $\alpha_1 \approx -0.468$ given in \eqref{eq:bb9_+bolt_alpha_1}.
The analysis of the extremal curve in the complex plus-Bolt geometry is more involved, but this time it will yield new constraints.
When $-1 < \alpha < 5 - 3\sqrt{3} \approx -0.196$ so that the square root in \eqref{eq:bb9_bolt_pm} is real, the sign of $\arg p_0$ depends on $\alpha$.
However, as in the minus-Bolt case, the extremal curve exits the region with $\arg p < 0$ at a point where $\Re \gamma_\mathsf{e} = \mathcal{O}(P^{-3})$ and $\Im \gamma_\mathsf{e} = \mathcal{O}(P^{-3/2})$ and we can use \eqref{eq:bb9_bolt_gamma_1} starting from the origin without a significant error.

\begin{figure}[ht!]
  \centering
  \subfloat[minus-Bolt]{\includegraphics[width=230pt]{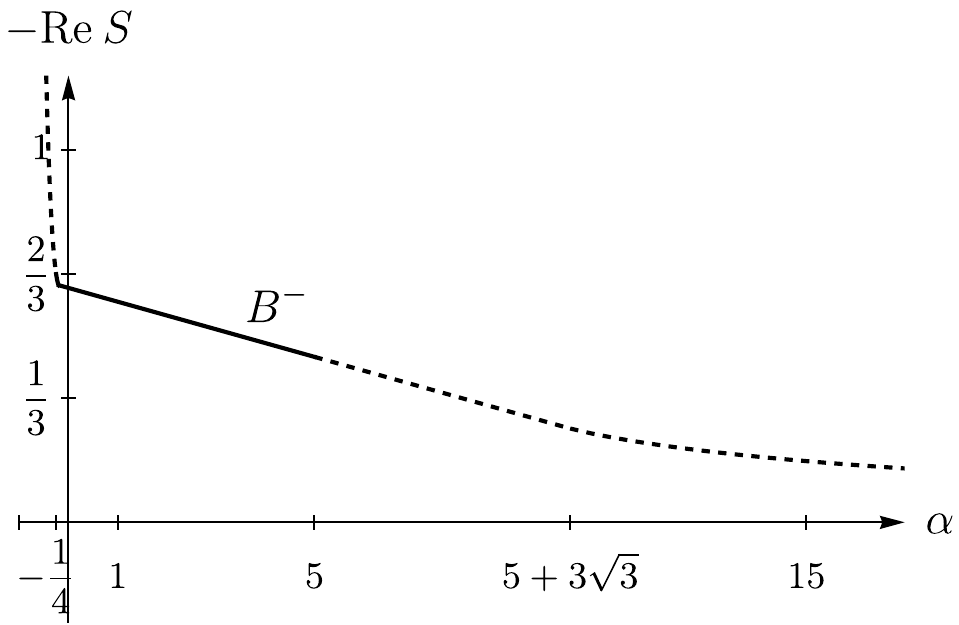}}
  \hfill
  \subfloat[plus-Bolt]{\includegraphics[width=230pt]{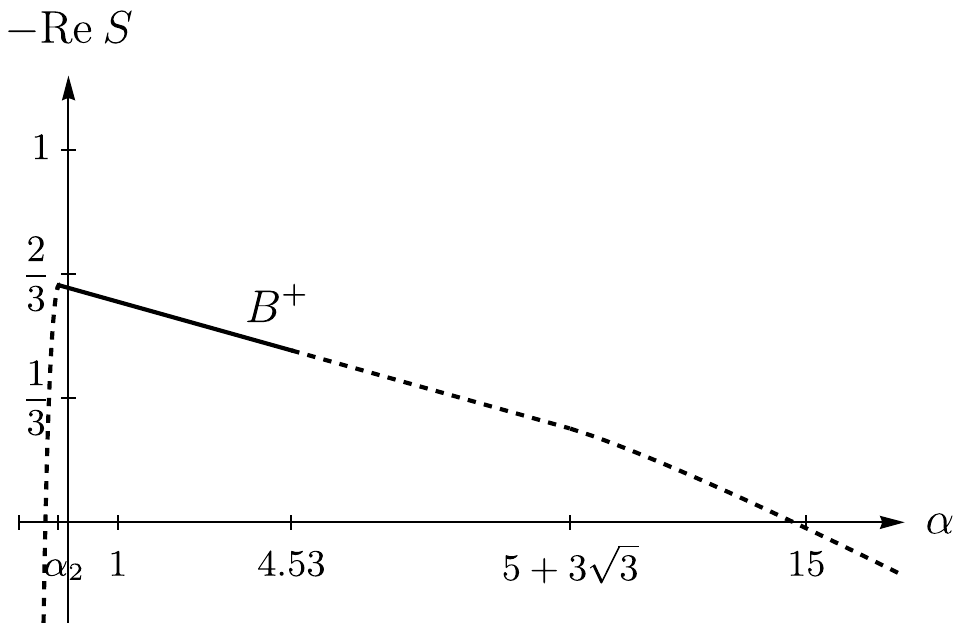}}
  \caption{The KSW criterion applied to the no-boundary state excludes large parts of the Taub-Bolt-dS branches in biaxial Bianchi IX minisuperspace.
    Shown is the real part of the action (in units of $4\pi^2 \Mpl^2/H^2$) of Bolt-type no-boundary saddles with a single-squashed $S^3$ boundary, in the large volume limit.
    Both Bolts have equal semiclassical weighting for $5 - 3\sqrt{3} < \alpha < 5 + 3\sqrt{3}$, although the solutions differ.
    The minus-Bolt is KSW-allowable (solid lines) for $-1/4 < \alpha < 5$ and the plus-Bolt is allowable for $\alpha_2 < \alpha < \alpha_3$, with $\alpha_2$ defined in \eqref{eq:bb9_bolt_alpha2} and $\alpha_3 \approx 4.530$ determined numerically.}
  \label{fig:bb9_Bolt_results}
\end{figure}

When $5 - 3 \sqrt{3} < \alpha < 5$ so that the square root in \eqref{eq:bb9_bolt_pm} is imaginary and there is an allowable direction $\gamma'$ at the origin, $\arg p_0 < 0$ by an $\mathcal{O}(P^0)$ amount and we have to use \eqref{eq:bb9_bolt_gamma_2} starting from the origin.
In this case, there are several lines on which $p \in \mathbb{R}$ or $q \in \mathbb{R}$ and regions where $2 \abs{\arg p} + \abs{\arg q} > \pi$ that $\gamma_\mathsf{e}$ risks intersecting, see Fig.~\ref{fig:bb9_Bolt_examples}.
However, these potential problems only occur in an $\mathcal{O}(P^0)$ neighborhood of the origin.
One can show, using the methods of Appendix~\ref{app:bb9_NUT_extremal_curve_validity}, that the extremal curve starting from any $\mathcal{O}(P^0)$ point $r$ with $0 < \Re r < 1/3$ and $\Im r > 0$ remains valid and passes to the left of $v$ as long as it avoids these problems close to the origin.
To determine allowability, we can thus check whether the extremal curve starting from the origin reaches $\Re \gamma_\mathsf{e} = 1/3$ before $\arg p > 0$ and $\arg q > 0$ (not allowable), or whether it reaches the region with $\arg p > 0$ and $\arg q > 0$ close to the vertical through $v$ and above all problematic regions (allowable).
In practice, we do this numerically using the strict $P \to \infty$ solution and check that $\gamma_\mathsf{e}$ remains to the left of $v$ at least until $\Im \gamma_\mathsf{e} = 3$.
With these methods, we find that
\begin{itemize}
  \item $-1 < \alpha < \alpha_1$ with $\alpha_1 \approx -0.468$ defined in \eqref{eq:bb9_+bolt_alpha_1}: the plus-Bolt violates the integrated 0-form criterion \eqref{eq:bb9_KS0_integrated} and is hence not allowable.
        Relatedly, the naive extremal curve passes to the right of the endpoint $v$.

  \item $\alpha_1 < \alpha < \alpha_2$ with
        \begin{equation}
          \label{eq:bb9_bolt_alpha2}
          15 \alpha_2^4 + 340 \alpha_2^3 - 267 \alpha_2^2 + 126 \alpha_2 + 43 = 0,
          \qquad \alpha_2 \approx -0.216 \+\colon
        \end{equation}
        The extremal curve crosses a line where $q \in \mathbb{R}$ at an $\mathcal{O}(P^0)$ distance from the origin.
        In this regime we resort to numerics and find strong evidence that the plus-Bolt is never allowable, see Fig.~\ref{fig:numerics}.

  \item $\alpha_2 < \alpha < 5 - 3 \sqrt{3} \approx -0.196$: the naive extremal curve is valid and passes to the left of $v$.
        We conclude that the plus-Bolt is allowable.

  \item $5 - 3 \sqrt{3} < \alpha < \alpha_3$ with $\alpha_3 \approx 4.530$ determined numerically: the extremal curve escapes the region where $\arg p < 0$ or $\arg q < 0$ close to the origin and continues to pass to the left of the endpoint $v$.
        The plus-Bolt is allowable.

  \item $\alpha_3 < \alpha < 5$: the extremal curve reaches $\Re \gamma_\mathsf{e} = \Re v = 1/3 + \mathcal{O}(P^{-1})$ at an $\mathcal{O}(P^0)$ distance from the origin.
        Hence, the plus-Bolt is not allowable.

  \item $5 < \alpha$: the plus-Bolt is not allowable since there is no allowable direction $\gamma'$ at the origin.
\end{itemize}
The transition at $\alpha = \alpha_3$ is illustrated in Fig.~\ref{fig:bb9_Bolt_examples}.
In summary, the KSW criterion applied to the no-boundary state in biaxial Bianchi IX minisuperspace also excludes a significant part of the Taub-Bolt-dS branch.
The results, for both Bolts, are summarized in Fig.~\ref{fig:bb9_Bolt_results}.

\subsection{KSW constraints on double-squashed Taub-NUT-dS} \label{doublesquashedsec} \noindent
In this section we return to the full double-squashed cosmologies \eqref{ds2double} where the second squashing $\beta \neq 0$.
In this case, as far as we are aware, analytic solutions to the classical equations of motion are not available and one must resort to numerical techniques.

Here, we do not aim for a complete treatment of allowability in the large volume limit ($a \to \infty$, $\alpha, \beta \to \text{constants}$) as we did above for the biaxial model.
Instead, we begin by recalling that KSW-allowable no-boundary solutions have an action with negative real part, $\Re S < 0$, as discussed in \S\ref{introsec}.
This conclusion does not require a detailed description of the solutions; it suffices to argue that the GHY term does not contribute to $\Re S$ in the large volume limit, as is the case for vacuum solutions that tend to empty de Sitter space.

\begin{figure}[ht!]
  \centering
  \subfloat[The region where $R^{(3)} < 0$ (black, striped)\hspace*{1em}]{\includegraphics[height=225pt]{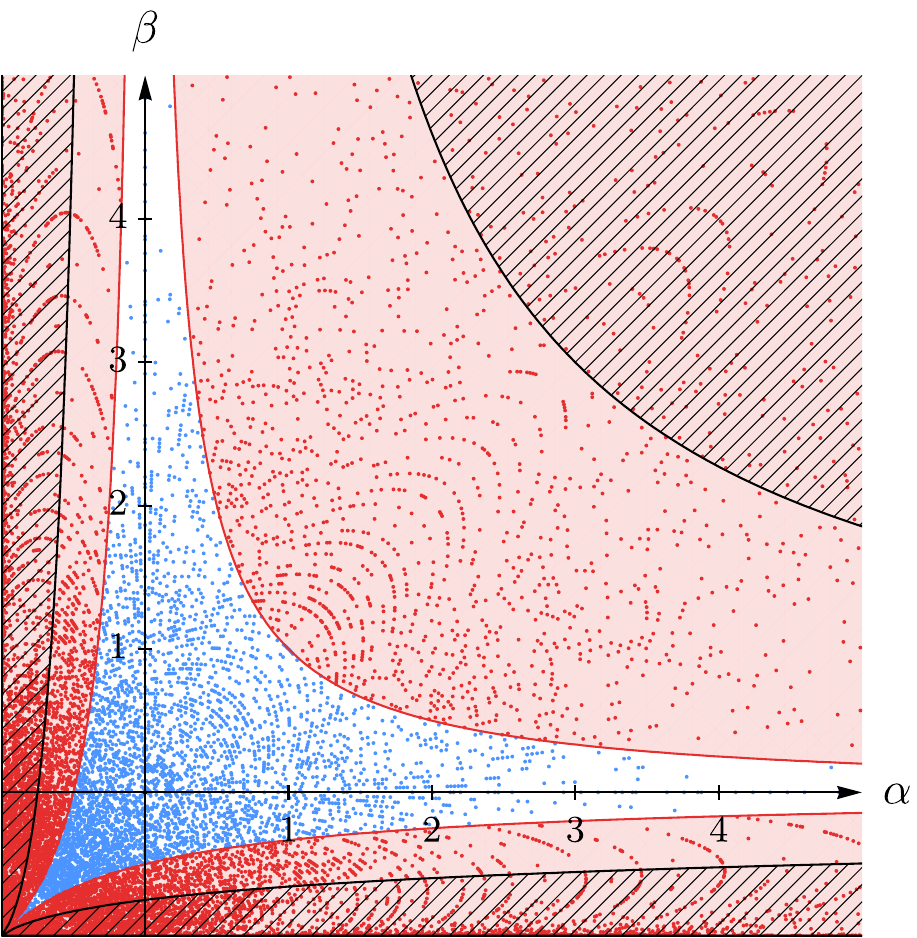}}
  \hfill
  \subfloat[Evidence for the conjecture at large $\alpha$]{\includegraphics[height=225pt]{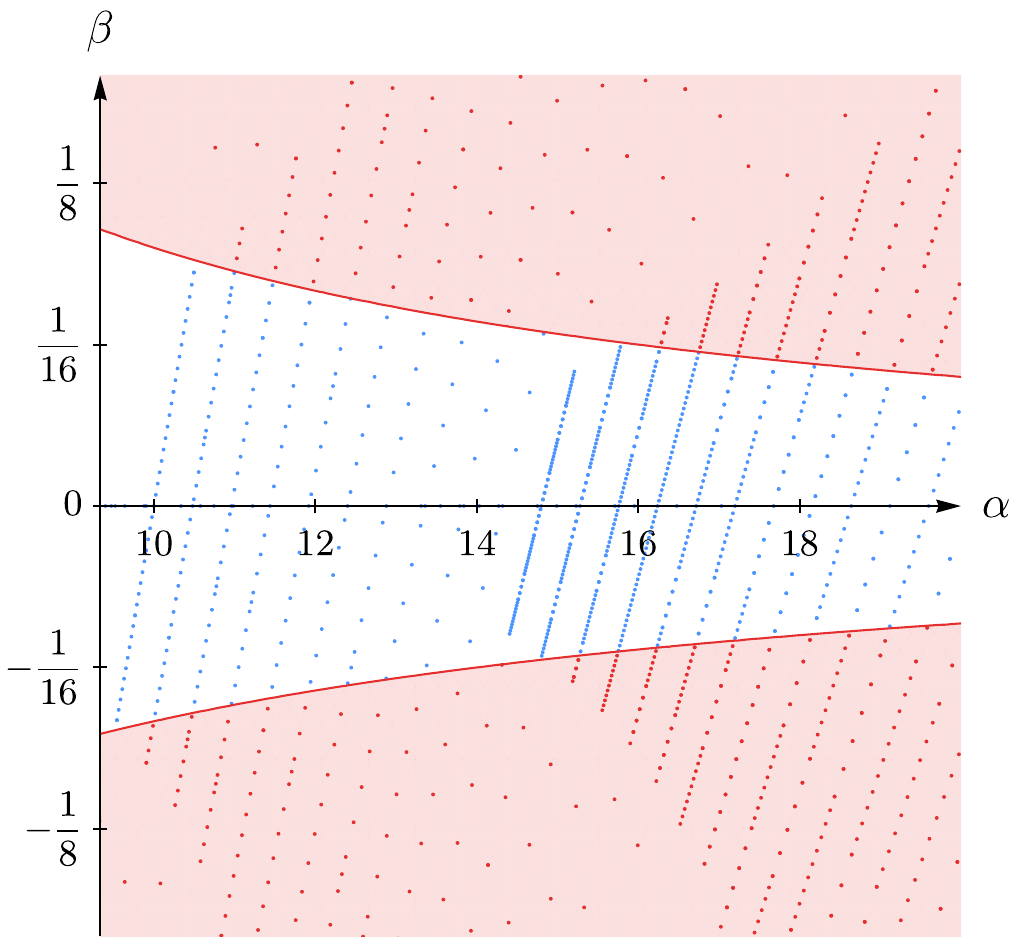}}
  \caption{The conjectured region \eqref{eq:double-squashed_conjecture} where $\Re S > 0$ (red), ruled out by KSW, and numerical solutions from \cite{Bobev:2016sap} (red/blue points for positive/negative $\Re S$) for infinite-volume NUT-type no-boundary solutions with double-squashed $S^3$ boundaries parametrized by $\alpha$ and $\beta$ in \eqref{ds2double}.
    All numerical solutions satisfy the conjecture.
    The region where $R^{(3)} < 0$ is contained in the region where $\Re S > 0$ and, hence, nonallowable.
    Note that there is a symmetry interchanging $\alpha$ and $\beta$.
  }
  \label{fig:double_squashed_NUT}
\end{figure}

In the single-squashed discussion of the preceding sections, for both NUT- and Bolt-type solutions, we noticed that the allowability constraint $\Re S < 0$ excludes the regime where the boundary Ricci curvature is negative.
For the double-squashed $S^3$ boundary at constant $t$ in \eqref{ds2double}, we have
\begin{equation} \label{R3boundary}
  R^{(3)} = \frac{6 + 8(\alpha + \beta) + 2\alpha\beta(6-\alpha\beta)}{a^2 (1+\alpha)(1+\beta)} \+,
\end{equation}
and one may wonder whether the conclusion extends, that is, whether the regime where $R^{(3)} < 0$ is contained within the regime where $\Re S > 0$, so that also here configurations with negative curvature $R^{(3)}$ would be nonallowable.
For the double-squashed NUT-type solutions ($a, \alpha, \beta \to 0$ at the south pole) described in \cite{Bobev:2016sap}, we find numerical evidence that this is the case.
Concretely, we conjecture that $\Re S > 0$ when
\begin{equation}
  \label{eq:double-squashed_conjecture}
  \alpha \beta > 1 \+,\quad
  (\alpha - \beta - 2)^2 - (\alpha + \beta + 2)^2 > 4 \quad\text{or}\quad
  (\beta - \alpha - 2)^2 - (\beta + \alpha + 2)^2 > 4 \+.
\end{equation}
We have verified this on the numerical double-squashed NUT-type solutions of \cite{Bobev:2016sap} and this also reduces to the correct constraint when $\alpha = 0$ or $\beta = 0$.
Assuming the above conjecture, the $R^{(3)} < 0$ region is contained within the $\Re S > 0$ region, as illustrated in Fig.~\ref{fig:double_squashed_NUT}, and hence excluded by KSW.
The above is only a necessary condition to satisfy KSW, presumably the full criterion rules out even more of the configuration space of double-squashed boundaries.%
\footnote{Indeed in \S\ref{sec:bb9_NUT_KS} we showed that on the $\beta = 0$ slice, only $\alpha > -1/6$ is KSW-allowable while $\alpha > -1/2$ has $\Re S < 0$.
  There is an $\alpha \leftrightarrow \beta$ symmetry, so the same is true on the $\alpha = 0$ slice.}
We leave this as an open question for future investigation.

Interestingly, this result is in line with dS/CFT, where the Ricci scalar on the conformal boundary enters as a ``mass'' term of conformally coupled scalars in the dual theory.
This suggests that regions of superspace corresponding to negative curvature configurations are dual to unstable boundary theories, with a divergent partition function \cite{Hawking:2017wrd}.

\section{\texorpdfstring{$S^1 \times S^2$}{S1 x S2} boundary} \label{S1S2sec} \noindent
In the second minisuperspace we consider, known as the Kantowski-Sachs model, the spatial slices have $S^1 \times S^2$ topology.
We write the 4D metric as
\begin{equation} \label{S1S2metric}
  H^2\+ \di s^2 = \frac{\di r^2}{a(r)^2} + a(r)^2 \di \omega^2 + b(r)^2 \di \Omega_2^2 \+,
\end{equation}
where $\omega \cong \omega + 2\pi$ and $\di \Omega_2^2$ is the round metric on the unit $S^2$.
We will denote the final values by $a(v) = A$, $b(v) = B$ and define the ratio
\begin{equation}
  \lambdabarr \equiv \frac{A}{B}
\end{equation}
of the size of the $S^1$ to the size of the $S^2$.
As with the parameters $\alpha$ and $\beta$ of the squashed $S^3$ minisuperspace considered in \S\ref{S3sec}, classical solutions are characterized by a constant value $\lambdabarr$ at large volume.
We therefore implement the large volume limit by taking $B \to \infty$ while keeping $\lambdabarr$ fixed.

This model has been studied by many authors, including \cite{LaflammeThesis,Anninos:2012ft,Conti:2014uda,Maldacena:2019cbz,Fanaras:2021awm}.
In particular, \cite{Maldacena:2019cbz} noticed that in the $\lambdabarr \to \infty$ limit, solutions \eqref{S1S2metric} of the Einstein equation, which we will discuss below, develop a long $(\text{``nearly''}\ \mathrm{dS}_2) \times S^2$ throat.
Fluctuations of the size of the $S^2$ can be identified with a dilaton in the $\mathrm{dS}_2$ factor.
More precisely, the nearly $\mathrm{dS}_2$ region is described by the de Sitter version of 2D Jackiw-Teitelboim gravity.
This theory has no propagating bulk degrees of freedom, but there is a fluctuating mode of the boundary $S^1$, which is described by a Schwarzian action and can be integrated out exactly at one loop \cite{Stanford:2017thb}.
This leads to a polynomially decaying prefactor, proportional to $\lambdabarr^{-\gamma}$ with $\gamma > 0$, in the wave function of the universe as $\lambdabarr \to \infty$.%
\footnote{More precisely, for $1 \ll \lambdabarr \lesssim \Mpl/H$, $\gamma \approx 3$ while for $\lambdabarr \gtrsim \Mpl/H$, $\gamma \approx 3/2$ \cite{Maldacena:2019cbz,Iliesiu:2020qvm}.}
Using the Klein-Gordon norm, the measure is $\di \lambdabarr$, and hence, the normalizability of the wave function is determined by an integral%
\footnote{The derivative in the Klein-Gordon norm contributes as a constant factor in the $\lambdabarr \gg 1$ regime.}
\begin{equation} \label{psinorm}
  \int \di \lambdabarr \, \frac{1}{\lambdabarr^{2\gamma}} \exp [- 2 \Re S(\lambdabarr) ] \+,
\end{equation}
where $S$ is the action of the dominant contributing saddle.
This form of the integrand is only valid at large $\lambdabarr \gg 1$, but this is sufficient to argue for the convergence of \eqref{psinorm} \cite{Maldacena:2019cbz} and thus for the normalizability of $\Psi_\textsf{HH}$.

We will return to \eqref{psinorm} below, first passing to a description of solutions of the form \eqref{S1S2metric} to the Einstein equation.
In terms of the variable $c(r) = a(r)^2 b(r)$, the action \eqref{theaction} reads
\begin{equation}
  S = \frac{8 \pi^2 \Mpl^2}{H^2} \biggl[ \int \di r \bigl( -b'c' + 3 b^2 - 1 \bigr) - \frac{1}{2} \bigl( 3 b'c+b c' \bigr)\big|_{r=0} \biggr] \+.
\end{equation}
The equations of motion read
\begin{equation}
  \label{eq:S1S2_eoms}
  b'' = 0 \+, \qquad
  b'c'+3b^2-1 = 0 \+,
\end{equation}
with general solution
\begin{equation}
  \label{eq:S1S2_general_solution}
  b(r) = b_0 + b_1 r \+,\qquad
  c(r) = c_0 + c_1 r + c_2 r^2 + c_3 r^3 \+,
\end{equation}
where $c_1 = (1-3b_0^2)/b_1$, $c_2 = -3b_0$ and $c_3 = -b_1$.
Imposing the final boundary conditions, we find $b_1 = (B-b_0)/v$, $c_0 = A^2 B-c_1 v-c_2 v^2-c_3 v^3$.
At this stage the remaining unknowns are $b_0$ and $v$. \\ \indent We must now choose how to smoothly ``cap off'' the geometry in the past to obtain a no-boundary solution.
This can be done by letting $a \to 0$ or $b \to 0$ as $r \to 0$.
The latter choice gives rise to a purely oscillating wave function \cite{LaflammeThesis} and so does not satisfy KSW by the argument of \S\ref{introsec} ($\Re K = 0$ and $v \in i\mathbb{R}$ here).
Proceeding with the former choice we set $c_0 = 0$, and the topology becomes that of $B^2 \times S^2$ provided $c_1^2 = 4 b_0^2$.
As before we may choose a sign without loss of generality; $c_1 = 2 b_0$.
At this stage we have eliminated all unknowns, but as for the Bolts of \S\ref{sec:bb9_Bolt_KS}, we again find two branches of solutions.
They can be identified by solving for $b_0$ as a function of $v$,
\begin{equation}
  b_0^{\pm}(v) = \frac{B \pm \sqrt{B^2-2v+3v^2}}{2-3v} \+,
\end{equation}
and we call these two branches $L^\pm$.
Since the situation is analogous to the Bolt calculation, we omit some details.
There is a communal polynomial equation for $v$,
\begin{equation}
  (3 B^2 - 4)v^5 - 2(3 B^2 - 4)v^4 + 2(3 \lambdabarr^2 B^4 + 2 B^2 - 2)v^3 - 8 \lambdabarr^2 B^4 v^2 + \lambdabarr^2 B^4 (3 \lambdabarr^2 B^2 + 4) v - 2 \lambdabarr^4 B^6 = 0 \+,
\end{equation}
with the following two relevant solutions at large $B$:%
\footnote{Apart from the complex conjugate solutions, there is a real solution $v = 2/3 - 8/(27 \lambdabarr^2 B^2) + \mathcal{O}(B^{-6})$.
  The corresponding saddle is real Euclidean, and hence KSW-allowable, with on-shell action $S = 27 \lambdabarr^4 B^6/16 + \mathcal{O}(B^2)$.
  Thus, it is exponentially suppressed with vanishing weight at infinite volume and we do not consider it further.}
\begin{equation} \label{vexpansionL}
  v^\pm = i \lambdabarr \left( B + \frac{1}{2B} \right) + \frac{1}{3} \left( 1 \pm \sqrt{1-3\lambdabarr^2} \right) + \mathcal{O}(B^{-2}) \quad \text{as } B \rightarrow \infty \+.
\end{equation}
As $\lambdabarr$ passes $1/\sqrt{3}$ from below, we will let $\sqrt{1-3\lambdabarr^2} \to i \sqrt{3\lambdabarr^2-1}$.
With this the two branches $L^\pm$ are well-defined for all $\lambdabarr > 0$.
The on-shell action evaluates to
\begin{equation}
  S^\pm = \frac{8 \pi^2 \Mpl^2}{H^2} \left\{ i \left( 2 \lambdabarr B^3 - \lambdabarr B \right) - \frac{1}{27} \left[ 9 \pm 6\sqrt{1-3\lambdabarr^2} - \frac{2}{\lambdabarr^2} \left( 1 \pm \sqrt{1-3\lambdabarr^2} \right) \right] + \mathcal{O}(B^{-1})\right\}
\end{equation}
in the large $B$ regime.
Hence, for $\lambdabarr > 1/\sqrt{3}$, the square root terms do not contribute to $\Re S$.

The solutions \eqref{S1S2metric} we have described are complexified versions of the Schwarzschild-de Sitter geometry.
This can be seen by the coordinate transformation $b(r) = \rho$, $\omega = i b_1 \tau$ in which the metric reads
\begin{equation} \label{SdSmetric}
  \begin{aligned}
    H^2\+ \di s^2 & = -f(\rho) \+ \di \tau^2 + \frac{\di \rho^2}{f(\rho)} + \rho^2 \di \Omega_2^2 \+, \\
    f(\rho)       & = 1 - \rho^2 - \frac{\mu}{\rho} \+, \quad \quad \mu = b_0 - b_0^3 \+.
  \end{aligned}
\end{equation}
The mass and inverse temperature of the black hole are
\begin{equation}
  M = 4 \pi \frac{\Mpl^2}{H} \mu \+, \quad \quad \beta_\textsf{BH} = \frac{4 \pi \rho_+}{1-3\rho_+^2} H^{-1} \+,
\end{equation}
where $\rho_+ = b_0$ is the location of the black hole horizon.
In the no-boundary solution, $b_0 \in \mathbb{C}$ and so both $M$ and $\beta_\textsf{BH}$ are complex.
In the classical regime $B \to \infty, \lambdabarr$ fixed, they are
\begin{equation}
  \begin{aligned}
    \mu               & = \frac{2i}{27 \lambdabarr^3} \left( 1 \pm \sqrt{1-3\lambdabarr^2} \right) \left( 1 + 3 \lambdabarr^2 \pm \sqrt{1-3\lambdabarr^2} \right) + \mathcal{O}(B^{-2}) \+, \\
    \beta_\textsf{BH} & = 2 \pi i \lambdabarr / H + \mathcal{O}(B^{-2}) \+,
  \end{aligned}
\end{equation}
where the two sign options correspond to the $L^\pm$ branches discussed above.
We see that the inverse temperature is purely imaginary and proportional to $\lambdabarr$.
We can go to Euclidean radius by $\rho = i \tilde{\rho}$,%
\footnote{Note that $\rho$, which is the radial coordinate of the black hole, plays the role of the time coordinate in the no-boundary solutions.}
in which case \eqref{SdSmetric} becomes minus the metric of a Euclidean AdS-Schwarzschild black hole with mass parameter $\tilde{\mu} = -i \mu$.
In this case the period of the thermal circle, or inverse temperature, is $-i \beta_\textsf{BH} = 2 \pi \lambdabarr / H$ for both types of solutions.
The $L^-$ solution has $\tilde\mu \sim \lambdabarr/2$ as $\lambdabarr \to 0$ (or $T_\textsf{BH} \to \infty$) while for the $L^+$ solution $\tilde\mu \sim 8/(27 \lambdabarr^3)$ in this limit, so it is the former that corresponds to the usual small AdS black holes in the high temperature regime.

We also have
\begin{align} \label{b0pmeq}
  \sqrt{3} \rho_+ = \sqrt{3} b_0^\pm = \frac{i}{\sqrt{3} \lambdabarr} \pm i \sqrt{\frac{1}{3 \lambdabarr^2}-1} + \mathcal{O}(B^{-2}) \to \frac{i}{\sqrt{3} \lambdabarr} \mp \sqrt{1 - \frac{1}{3 \lambdabarr^2}} + \mathcal{O}(B^{-2}) \+,
\end{align}
where left of the arrow is for $\lambdabarr < 1/\sqrt{3}$ and right of the arrow is for $\lambdabarr > 1/\sqrt{3}$.
This allows for a comparison with \cite{Maldacena:2019cbz}, where they chose the $L^-$ branch only to contribute to the wave function, assuming the $L^+$ branch is not picked up by the contour of integration in the quantum gravity path integral.
Instead, here we will let the KSW criterion decide which saddles contribute to the semiclassical no-boundary wave function.

\subsection{KSW constraints on Kantowski-Sachs} \label{KantSsec} \noindent
We now examine the allowability of the solutions $L^\pm$ described above, setting $r = \gamma(\ell)$ in \eqref{S1S2metric} as we did previously.
A simple constraint on both solutions follows from the asymptotics \eqref{b0pmeq}.
Considering the metric \eqref{S1S2metric} near the origin where the $S^2$ has finite size $b_0$, we find from \eqref{b0pmeq} that $| \arg (b_0^\pm)^2 | > \pi/2$ for $\lambdabarr < \sqrt{2/3}$, and hence such values cannot be KSW-allowable.

For the $L^-$ solution, we claim that the condition $\lambdabarr > \sqrt{2/3}$ covers all KSW constraints.
To see this, note that near the origin, $a(r)^2 = 2r + \mathcal{O}(r^2)$.
Hence, in this neighborhood the extremal curve equation becomes, since $\arg (b_0^-)^2 > 0$,
\begin{equation} \label{eq:S1S2_KSW}
  \biggl|\arg \frac{\gamma'^{\+ 2}}{a(\gamma)^2}\biggr| + \abs{\arg a(\gamma)^2} + 2 \abs{\arg b(\gamma)^2} = 2 \arg \gamma' + 2 \arg b(\gamma)^2 = \pi \+.
\end{equation}
One may solve this equation by integration, similarly to what was done in \S\ref{S3sec}.
We find
\begin{equation}
  \Re \left( b_0 \gamma_\mathsf{e} b(\gamma_\mathsf{e}) + \frac{b_1^2 \gamma_\mathsf{e}^3}{3} \right) = 0 \+.
\end{equation}
It can readily be shown that this equation with $\gamma_\mathsf{e} \to v$ has no solutions for any $\lambdabarr > \sqrt{2/3}$.
This implies that if the naive extremal curve -- the one obtained by omitting the absolute values, as we have done -- is the true extremal curve, in other words, if it does not cross a line where $a^2 \in \mathbb{R}$, $b^2 \in \mathbb{R}$ or $b^2 \sqrt{a^2} \in i\mathbb{R}$, then all such values of $\lambdabarr$ are allowable.
There are no such crossings, it turns out.
The line where $b^2 \in \mathbb{R}$ that $\gamma_\mathsf{e}$ might intersect is the vertical $\Re r = \Re v$, up to $1/B$ corrections that give it a positive slope and make an intersection impossible since $\gamma' = i/b(\gamma)$.
The curves where $a^2 \in \mathbb{R}$ lie outside of the rectangle with corners in $r = 0$ and $r = v$ and are, thus, not a problem.
Lastly, even though there is only a small angular opening of allowable directions at the origin when $\lambdabarr$ is close to, but greater than, $\sqrt{2/3}$, the naive extremal curve does not cross $b^2 \sqrt{a^2} \in i\mathbb{R}$.

For the $L^+$ solution we claim \emph{no} values of $\lambdabarr$ are KSW-allowable.
One way to see this is by considering the horizontal line $r = x + i \sqrt{3\lambdabarr^2-1}/3$.
On this line we have
\begin{equation}
  b(r)^2 = - \frac{(3x-1)^2}{9 \lambdabarr^2} + \mathcal{O}(B^{-2}) \+ .
\end{equation}
That is, $b(r)^2$ is real nonpositive.
An allowable curve may not cross this line, avoiding the point where $b=0$ to not have a singular solution, because there
\begin{equation}
  \biggl|\arg \frac{\gamma'^{\+ 2}}{a(\gamma)^2}\biggr| + \abs{\arg a(\gamma)^2} + 2 \abs{\arg b(\gamma)^2} \geq 2 \abs{\arg b(\gamma)^2} > \pi \+.
\end{equation}

\begin{figure}[ht!]
  \centering
  \subfloat[]{\includegraphics[width=230pt]{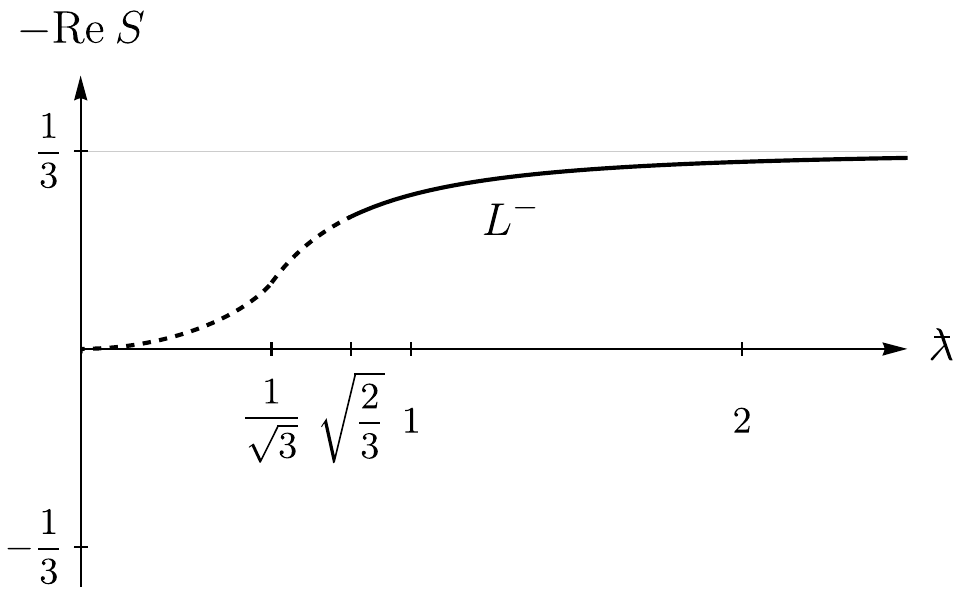}}
  \hfill
  \subfloat[]{\includegraphics[width=230pt]{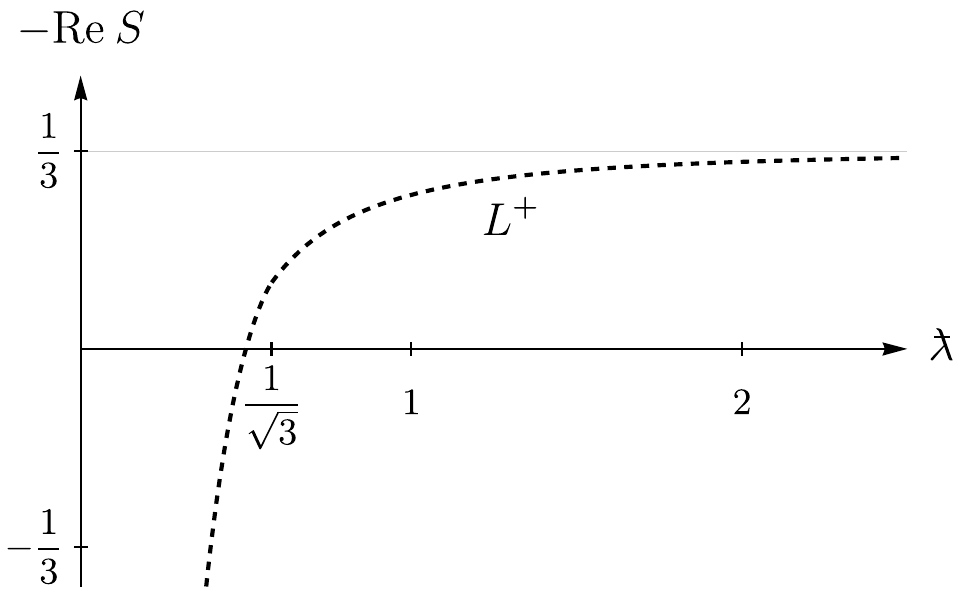}}
  \caption{The KSW criterion applied to the no-boundary state strongly constrains the configuration space in the Kantowski-Sachs minisuperspace.
    Shown are the real parts of the Euclidean action (in units of $8\pi^2 \Mpl^2/H^2$) of the Kantowski-Sachs no-boundary saddles $L^\pm$, with $S^1 \times S^2$ boundary, at large volume.
    In this limit, $\Re S$ depends only on the ratio $\lambdabarr$ of the radii of the $S^1$ and $S^2$ factors.
    The branches $L^\pm$ yield the same semiclassical weighting for $\lambdabarr > 1/\sqrt{3}$, which asymptotes to $1/3$ as $\lambdabarr \to \infty$, but the solutions differ as complexified geometries nevertheless.
    The KSW criterion excludes the $L^+$ branch entirely but allows $L^-$ saddles for sufficiently low temperatures $\beta_\textsf{EAdS} H/(2\pi) = \lambdabarr > \sqrt{2/3}$ (solid line).}
  \label{fig:S1xS2_results}
\end{figure}

In conclusion, the KSW allowability criterion strongly constrains the configuration space in the Kantowski-Sachs ($S^1 \times S^2$) minisuperspace, excluding in particular the entire high temperature regime where the $S^1$ is sufficiently small compared to the $S^2$.
Fig.~\ref{fig:S1xS2_results} depicts these constraints for both branches of saddles.

This result is consistent with the choice made in \cite{Maldacena:2019cbz}, except that here the part with $\lambdabarr < \sqrt{2/3}$ is further excised from $\Psi_\textsf{HH}$.%
\footnote{Presumably there exists an allowable four-geometry with less symmetry that contributes in an exponentially suppressed way to $\Psi_\textsf{HH}$ in this regime.}
As discussed in \cite{Maldacena:2019cbz} and above, the normalizability of $\Psi_\textsf{HH}$ depends on the convergence of the integral \eqref{psinorm}.
For both solutions, we have
\begin{equation}
  \Re S (\lambdabarr) = \text{constant} + \frac{16 \pi^2}{27} \frac{\Mpl^2}{H^2} \frac{1}{\lambdabarr^2} \quad \text{for } \lambdabarr \geq \frac{1}{\sqrt{3}} \+.
  \label{largelambdaS}
\end{equation}
Thus, the convergence of the integral as $\lambdabarr \to \infty$ is ensured by the polynomially decaying prefactor.
Further, because of the large factor $\Mpl^2/H^2 \gg 1$ in \eqref{largelambdaS}, $\Psi_\textsf{HH}$ is peaked at large $\lambdabarr = \mathcal{O}(\Mpl/H)$ where the calculation of the prefactor is reliable, with an exponential damping for smaller values $1 \ll \lambdabarr \lesssim \Mpl/H$.
While this is a strong indication that the integral converges, one might worry about a divergence on the $\lambdabarr \to 0$ side of the integral where we do not have a handle on the prefactor.
The KSW analysis shows this part of the minisuperspace must be excised, providing further evidence of the normalizability of $\Psi_\textsf{HH}$.

\section{Conclusion} \label{sec:conclusions} \noindent
We have shown that the KSW criterion applied to the no-boundary state selects a specific subset of homogeneous but anisotropic deformations of de Sitter space.

In both Bianchi IX (squashed $S^3$) and Kantowski-Sachs ($S^1 \times S^2$) minisuperspace, KSW mostly excludes saddles that strongly deviate from isotropic de Sitter space.
The range of squashed surface configurations for which the corresponding saddles are not allowed by KSW includes all three-geometries with negative scalar curvature.
This applies not only to the Taub-NUT-dS and Taub-Bolt-dS branches of saddles in biaxial (single-squashed) Bianchi IX minisuperspace but also to Taub-NUT-dS saddles generating configurations that are double-squashed spheres.
The observation that KSW excludes negative curvature configurations resonates with dS/CFT reasoning, as we discussed in the Introduction.
It may also have implications for our understanding of eternal inflation.

Regarding the physical interpretation of \emph{why} KSW rules out certain saddles, in the Introduction we gave a simple argument of why the 0-form condition (i.e.~the mass of a probe scalar may not change sign during the evolution) selects those saddles with $\Re S < 0$.
This result assumes we are in the late-time classical regime of pure gravity with a positive cosmological constant.%
\footnote{In inflation, $\Re S \propto -\Mpl^4/V_*$ is always negative, yet, the 0-form condition may fail when backreaction is large \cite{Hertog:2023vot,Maldacena:2024uhs,Janssen:2024vjn}.}
This observation suffices to rule out the squashed $S^3$ boundaries with negative curvature discussed above.
As we have seen, the full KSW criterion may rule out more of the minisuperspace, however.
A clean example are the Taub-NUT-dS solutions with a single-squashed $S^3$ boundary discussed in \S\ref{sec:bb9_NUT_KS}.
For squashing parameters $\alpha < -1/2$, the 0-form condition fails since $\Re S > 0$, while for $\alpha > -1/2$ it is valid, i.e.~we can always find a representation of the no-boundary saddle in which the real part of the mass term $\sqrt{g} \+ m^2 \phi^2$ remains positive.
For $-1/2 < \alpha < -1/6$, however, we find that the higher-form criteria are necessarily violated in \emph{any} complex representation of the no-boundary saddle that satisfies the $0$-form criterion.%
\footnote{By representation we mean the choice of complexification of the time coordinate connecting the late-time boundary with the smooth ``beginning'', i.e.~the curve $\gamma$.} All KSW subcriteria are satisfied when $\alpha > -1/6$.

In Kantowski-Sachs minisuperspace we find that KSW selects the low-temperature regime of configuration space.
It would be interesting to investigate whether the critical value of the $S^1$ to $S^2$ ratio, $\lambdabarr = \sqrt{2/3}$, that signals the transition from allowed to excluded by KSW has an independent physical interpretation.

As an aside, we note that our analysis reveals a subtle but potentially important difference between the bulk formulation of the semiclassical no-boundary wave function \cite{Halliwell:2018ejl} and its holographic form \cite{Maldacena:2002vr,Hertog:2011ky}, in regions of superspace where more than one saddle potentially contributes.
The standard bulk expression $\Psi_\textsf{HH} \approx \sum_i e^{-S^{(i)}_\textsf{dS}}$ selects the saddle that minimizes $\Re S_\textsf{dS}$.
For each asymptotically de Sitter saddle, the on-shell action is related to that of an AdS domain wall by $\Re S_\textsf{dS} = - \Re S_\textsf{EAdS}$ \cite{Hertog:2011ky}.
Thus, the holographic expression $\Psi_\textsf{HH} \sim 1/Z$ selects the saddle that minimizes $\Re S_\textsf{EAdS}$ and hence \emph{maximizes} $\Re S_\textsf{dS}$.
Although not emphasized there, this can be seen from the results in \cite{Bobev:2016sap} where the Hawking-Page-type phase transition from the NUT to the plus-Bolt at $\alpha = 6 + 2\sqrt{10}$ is discussed from a holographic viewpoint.
The bulk formulation of the semiclassical no-boundary wave function instead predicts a transition from NUT to Bolts at $\alpha = 2$ and then back to NUT at $\alpha = 5$, after which neither Bolt is allowable by KSW, see Figs.~\ref{fig:bb9_NUT_results} and \ref{fig:bb9_Bolt_results}.%
\footnote{Without KSW, the minus-Bolt would be the dominant saddle at large $\alpha$ in the bulk formulation.}
For $\alpha < -1/6$ neither NUT nor Bolts are KSW-allowable; there may exist an allowable geometry of another type, but, presumably, its contribution is exponentially suppressed compared to the naive NUT/Bolt ones.

We note that in both minisuperspace models, the KSW criterion renders the semiclassical wave function ``nearly'' normalizable.
The only remaining possible divergences lie on the positive three-curvature side of the Taub-NUT-dS branch in Bianchi IX and in the low-temperature limit in the $S^1 \times S^2$ model, and are mild in the sense that one might expect these to be cured by the one-loop prefactors.
Indeed this was shown in \cite{Maldacena:2019cbz} to be the case for the $S^1 \times S^2$ model.

We conclude with an expression of excitement that the KSW criterion in conjunction with the no-boundary proposal leads to a non-trivial \emph{theoretical} prior on large-scale properties of our universe; we have shown here that certain anisotropic deformations of de Sitter are disfavored, while \cite{Hertog:2023vot,Maldacena:2024uhs,Janssen:2024vjn} showed that inflationary models with large $r$ are disfavored, two constraints which are in agreement with observations.
On the other hand, violations of the KSW criterion have been noted in the literature, e.g.~\cite{Chen:2023hra} (but see \cite{Chakravarty:2024bna}), which nevertheless yield physically sensible results.
Understanding when exactly the KSW criterion applies thus remains a pertinent question that we plan to investigate in the near future.

\section*{Acknowledgments} \noindent
We thank V.~Gorbenko, C.~Jonas, M.~Mirbabayi and Y.~Vreys for useful conversations.
T.~H.\ and J.~K.\ acknowledge support from the PRODEX grant LISA - BEL (PEA 4000131558), the FWO Research Project G0H9318N and the inter-university project iBOF/21/084.
J.~K.\ is also supported by the Research Foundation - Flanders (FWO) doctoral fellowship 1171823N.

\appendix
\section{Details of the extremal curve} \label{app:bb9_NUT_extremal_curve_validity} \noindent
In this appendix, we analyze the extremal curve $\gamma_\mathsf{e}$ for the Taub-NUT-dS solution.
The Bolt analysis is similar but slightly more involved technically due to lengthy expressions, and we only briefly comment on it at the end.
Recall that, as long as $0 < \arg p(\gamma_\mathsf{e})$, $0 < \arg q(\gamma_\mathsf{e})$ and $\arg q(\gamma_\mathsf{e}) + 2 \arg p(\gamma_\mathsf{e}) < \pi$, the extremal curve is given by \eqref{eq:bb9_NUT_extremal_curve}.
Below, we check when these assumptions are valid.

The real part of \eqref{eq:bb9_NUT_extremal_curve} is a parameterization-independent algebraic equation for $\gamma_\mathsf{e}$.
Writing $\gamma_\mathsf{e} = x + i y P^k$ with $x$, $y$ of order $P^0$, since we are interested in solutions to this equation with $0 \leq \Re \gamma_\mathsf{e} \leq \Re v$, it becomes
\begin{equation}
  \label{eq:bb9_NUT_large_P_extremal_curve}
  \begin{alignedat}[b]{2}
    0 & = - \frac{P^2}{3 (1 + \alpha)^2} \bigl[ x^2 \bigl((1+\alpha) x - 3\bigr) + \mathcal{O}(P^{-1})\bigr]
    + \mathcal{O}(P^{k + 1/2})
    \\ & \quad
       + \frac{P^{2k + 2}}{(1+\alpha)^3} \bigl[ (1+\alpha) \bigl((1+\alpha) x - 1\bigr) y^2 + 4 \alpha \bigl((1+\alpha) x - 2\bigr) y^2 P^{-1} + \mathcal{O}(P^{-2})\bigr]
    \\ & \quad
       - \frac{8 P^{3k + 1/2}}{3 (1+\alpha)^{5/2}} \bigl[\alpha y^3 + \mathcal{O}(P^{-1})\bigr] \+.
  \end{alignedat}
\end{equation}
We view this as a cubic equation for $y$ and solve it perturbatively in large $P$.
By examining when terms compete, we see that the solutions have either $k = 0$ or $k = 3/2$ when $x - (1 + \alpha)^{-1} = \mathcal{O}(P^0)$.
Since $\gamma_\mathsf{e}$ starts at the origin, it is clearly the $k = 0$ solution that is of interest.
Setting $k = 0$, \eqref{eq:bb9_NUT_large_P_extremal_curve} is solved to leading order in $1/P$ by
\begin{equation}
  \label{eq:bb9_NUT_extremal_curve_close_to_0}
  y = \pm x \sqrt{\frac{3 - (1+\alpha) x}{3 - 3 (1+\alpha) x}} \+,
\end{equation}
where the plus sign corresponds to the solution starting at an angle $\arg \gamma_\mathsf{e}'(0) = \pi /4$.
Again, this is only valid when $(1 + \alpha) x$ is not too close to one, i.e.~away from the vertical $x = \Re v$.

If $\gamma_\mathsf{e}$ intersects $p \in \mathbb{R}$ it has to be vertical at this point, since $\gamma_\mathsf{e}' = i/p$.
This never happens in the region analyzed so far, as is clear from \eqref{eq:bb9_NUT_extremal_curve_close_to_0}.

When $\Re \gamma_\mathsf{e}$ is close to $\Re v = (1 + \alpha)^{-1} + \mathcal{O}(P^{-1})$, a separate analysis is needed.
In this case, we write $\gamma_\mathsf{e} = (1+\alpha)^{-1} + (1+\alpha)^{-2} (8\alpha + \tilde{x}) P^{-1} + i y P^k$ so that $\tilde{x} = 0$ corresponds to $\Re \gamma_\mathsf{e} = \Re v$ up to $1/P$ corrections.
Now the $\mathcal{O}(P^{2k+2})$ term in \eqref{eq:bb9_NUT_large_P_extremal_curve} vanishes and all three solutions for $y$ are at $k = 1/2$.
Keeping also the $\mathcal{O}(P^{2k+1})$ term in \eqref{eq:bb9_NUT_large_P_extremal_curve}, we get the equation
\begin{equation}
  \label{eq:bb9_extremal_curve_close_to_vertical}
  0 = \bigl[2 (1+\alpha)^{-1/2} + (1+\alpha) y^2 \bigl(3 (1+\alpha)^{-1/2} (4\alpha + \tilde{x}) - 8 \alpha y\bigr)\bigr] P^2 + \mathcal{O}(P) \+.
\end{equation}
To find $y$ to leading order we drop the subleading $\mathcal{O}(P)$ term and get a cubic equation for $y$.
The structure of the real roots of this cubic depends on the sign of $\alpha$ and will tell us about intersections with $p \in \mathbb{R}$.
By looking at the extrema and asymptotics of the cubic we find the following:
For $\alpha > 0$, there is a single positive root for all values of $\tilde{x}$ and this root is larger than $(1+\alpha)^{-1/2}$ when $\tilde{x} = 0$, i.e.~$\gamma_\mathsf{e}$ never intersects $p \in \mathbb{R}$ and passes above the endpoint $v$.
For $\alpha < 0$, the cubic has a positive double root at $\tilde{x} = - 4\alpha - 2 (-2\alpha)^{2/3}$, two positive roots when $\tilde{x}$ is smaller and no positive root when $\tilde{x}$ is larger.
Hence, $\gamma_\mathsf{e}$ is vertical and intersects $p \in \mathbb{R}$ at the double root and our analysis is not valid past that point.
When $-1/2 < \alpha < 0$ ($-1 < \alpha < -1/2$), the point where $\gamma_\mathsf{e}$ intersects $p \in \mathbb{R}$ is above and to the left (below and to the right) of the endpoint $v$.
This analysis thus shows that $-1 < \alpha < -1/2$ is not allowable in the large $P$ limit, consistent with what we found by integrating the 0-form criterion (see footnote~\ref{footnote:bb9_NUT}).
As long as $\gamma_\mathsf{e}$ does not intersect $q \in \mathbb{R}$, which we have so far assumed, we conclude that $\alpha > -1/2$ is allowable in the large $P$ limit.
However, we will see below that this assumption is only valid for $\alpha > -1/6$.

Next, we check whether $\gamma_\mathsf{e}$ intersects $q \in \mathbb{R}$.
The equation $\Im q(r) = 0$ can be viewed as a relation between $\Re r$ and $\Im r$, similar to \eqref{eq:bb9_NUT_large_P_extremal_curve}.
Explicitly, writing $r = x + i y P^k$ with $x$ and $y$ assumed to be $\mathcal{O}(P^0)$,
\begin{equation}
  \label{eq:bb9_NUT_large_P_real_q}
  \begin{aligned}[b]
    0 & = \frac{16 P^{1/2}}{(1+\alpha)^{5/2}} \bigl[\alpha (2-x) x^2 + \mathcal{O}(P^{-1})\bigr]
    \\ & \quad
       - \frac{2 P^{k + 2}}{(1+\alpha)^2} \bigl[\xi(x) y + \mathcal{O}(P^{-1})\bigr]
    + \mathcal{O}(P^{2k+1/2})
    \\ & \quad
       - \frac{2 P^{3k+2}}{(1+\alpha)^2} \bigl[(1+\alpha) \bigl((1+\alpha)x - 1\bigr) y^3 - 4 \alpha y^3 P^{-1} + \mathcal{O}(P^{-2})\bigr] \+,
  \end{aligned}
\end{equation}
where $\xi(x) = (1+\alpha)^2 x^3 - 5 (1+\alpha) x^2 + 8 x - 4$.
Away from $x = 0$, $x = \Re v$ and $x$ such that $\xi(x) = 0$, solutions may exist for $k = -3/2$ and $k = 0$ when $\alpha > -1/2$ by an $\mathcal{O}(P^0)$ amount.
Since $\Im (\gamma_\mathsf{e}) = \mathcal{O}(P^0)$ in this region, only the $k = 0$ solutions are of interest.
Setting $k = 0$, we find the solutions $y = 0$ (which is the $k=-3/2$ solution in disguise) and%
\footnote{These solutions do not always exist for the values of $x$ of interest, i.e.~the $y$ given here can be imaginary.
  It is straightforward to check when this happens but we will not need it explicitly in what follows.}
\begin{equation}
  \label{eq:bb9_NUT_real_q_close_to_0}
  y = \pm \sqrt{\frac{\xi(x)}{(1+\alpha)\bigl(1 - (1+\alpha) x\bigr)}} \+,
\end{equation}
to leading order in $1/P$.
When $\xi(x) = 0$, all three roots of \eqref{eq:bb9_NUT_large_P_real_q} have $k=-1/2$ and there can be no intersection with $\gamma_\mathsf{e}$.

\begin{figure}[ht!]
  \centering
  \subfloat[$\alpha = -1/3$, $\Re v = 3/2$]{\includegraphics[width=230pt]{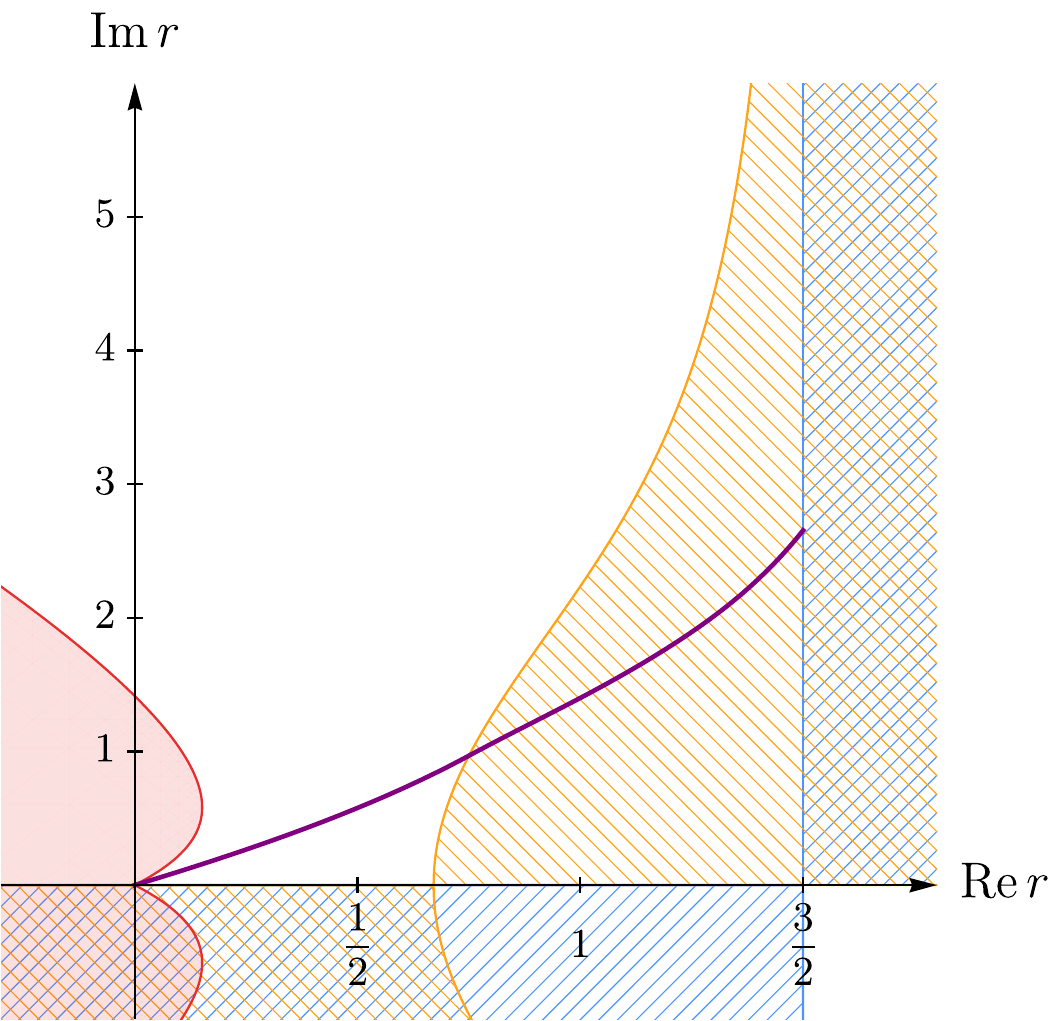}}
  \hfill
  \subfloat[$\alpha = -1/12$, $\Re v = 12/11$]{\includegraphics[width=230pt]{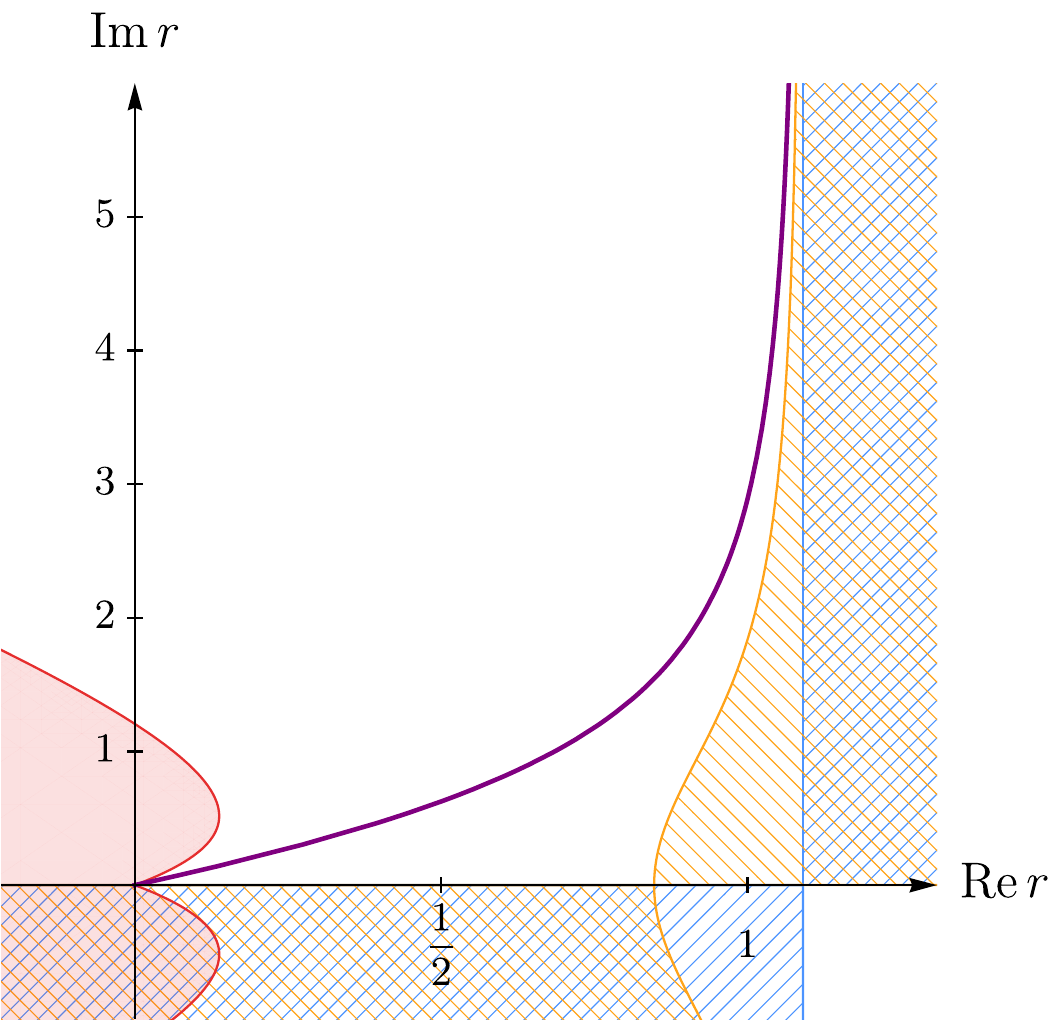}}
  \caption{Illustration of the different behaviors of the extremal curve (purple lines, obtained numerically) depending on whether $\alpha < -1/6$ or $\alpha > -1/6$ for Taub-NUT-dS solutions with $P \to \infty$.
    The colored regions indicate where $\arg p \leq 0$ (blue NE-SW-striped regions), $\arg q \leq 0$ (orange NW-SE-striped regions) and $2\abs{\arg p} + \abs{\arg q} \geq \pi$ (red regions), outside which the naive extremal curve is valid.
    Note that $\Re v$ is finite while $\Im v \to \infty$ when $P \to \infty$.
    The $\alpha = -1/3$ saddle is not allowable since the extremal curve reaches $\Re \gamma_\mathsf{e} = \Re v$ at a finite distance from the origin.}
  \label{fig:bb9_NUT_examples}
\end{figure}

Eqns.~\eqref{eq:bb9_NUT_extremal_curve_close_to_0} and \eqref{eq:bb9_NUT_real_q_close_to_0} describe the curves $\gamma_\mathsf{e}$ and $q \in \mathbb{R}$, respectively, by writing their imaginary parts $y$ as a function of their real parts $x$.
Hence, the two intersect where
\begin{equation}
  3 \xi(x) = (1 + \alpha) x^2 \bigl(3 - (1+\alpha) x\bigr) \+,
\end{equation}
i.e.~at the roots of a cubic polynomial.
By examining the extrema of this cubic and its values at $x = 0$ and $x = (1+\alpha)^{-1} = \Re v + \mathcal{O}(P^{-1})$, we conclude that there is a root in between these values precisely when $\alpha < -1/6$.
Thus, the curve defined by \eqref{eq:bb9_NUT_extremal_curve} becomes invalid at an $\mathcal{O}(P^0)$ distance from the origin when $\alpha < -1/6$ due to an intersection with $q \in \mathbb{R}$.
This transition is illustrated in Fig.~\ref{fig:bb9_NUT_examples}.

We also have to look for an intersection of $q \in \mathbb{R}$ and $\gamma_\mathsf{e}$ close to the vertical through $v$.
In this case, the $\mathcal{O}(P^{3k+2})$ term vanishes and the solutions are at $k=-3/2$ and $k=1/2$.
Since $\Im(\gamma_\mathsf{e}) = \mathcal{O}(P^{1/2})$ in this region, we focus on the $k = 1/2$ solutions.
Writing $r = (1+\alpha)^{-1} + (1+\alpha)^{-2} (8\alpha + \tilde{x}) P^{-1} + i y P^k$ as above and setting $k = 1/2$ the equation $\Im(q) = 0$ becomes
\begin{equation}
  0 = \frac{2 P^{5/2}}{(1 + \alpha)^3} \bigl[ \bigl(4\alpha - (1+\alpha)(4\alpha + \tilde{x}) y^2\bigr) y + \mathcal{O}(P^{-1})\bigr] \+.
\end{equation}
Dropping the subleading $\mathcal{O}(P^{-1})$ term and discarding the $y = 0$ solution (again, the $k=-3/2$ solution in disguise), we find that to leading order
\begin{equation}
  \label{eq:bb9_real_q_close_to_vertical}
  y = \pm \sqrt{\frac{4 \alpha}{(1+\alpha)(4\alpha +  \tilde{x})}} \+.
\end{equation}
For $\alpha > 0$, this root only exists for $ \tilde{x} > -4\alpha$ and then $y > (1+\alpha)^{-1/2}$ for $ \tilde{x} < 0$.
This means that, to the left of the vertical through $v$, the line $q \in \mathbb{R}$ that might intersect $\gamma_\mathsf{e}$ is above $v$ and any potential intersection does not affect the validity of $\gamma_\mathsf{e}$ in the region of interest.
Thus, we conclude that the Taub-NUT solution with fixed $\alpha > 0$ is allowable in the $P \to \infty$ limit.

The remaining region is when we are close to the vertical through $v$ and $-1/6 < \alpha < 0$.
Substituting the positive root in \eqref{eq:bb9_real_q_close_to_vertical} into \eqref{eq:bb9_extremal_curve_close_to_vertical} gives
\begin{equation}
  4 \alpha (1+\alpha) \biggl[\frac{4 \alpha}{(1+\alpha)(4\alpha + \tilde{x})}\biggr]^{3/2} = \frac{1 + 6\alpha}{\sqrt{1+\alpha}} \+.
\end{equation}
When $-1/6 < \alpha < 0$, the left-hand side is positive but the right-hand side negative for $\tilde{x}$ in the range such that $y$ in \eqref{eq:bb9_real_q_close_to_vertical} is real.
Thus, we conclude that the Taub-NUT solution with fixed $\alpha$ between $-1/6$ and $0$ is also allowable in the large $P$ limit.

Lastly, we check that $\arg q + 2 \arg p < \pi$ along $\gamma_\mathsf{e}$.
When $\gamma_\mathsf{e}$ is close to the vertical $\Re r = \Re v$ this inequality is satisfied since $p$ and $q$ are approximately real there.
This also implies that we can safely conclude that a saddle is allowable as long as the extremal curve passes to the left of $v$ since the horizontal distance between $\gamma_\mathsf{e}$ and $v$ is suppressed by $1/P$.
When $\gamma_\mathsf{e}$ is an $\mathcal{O}(P^0)$ distance from the origin, the inequality can be verified by looking for points along $\gamma_\mathsf{e}$ on which $p \sqrt{q} \in i\mathbb{R}$, or equivalently $p^2 q \in \mathbb{R}_-$.
By using \eqref{eq:bb9_NUT_extremal_curve_close_to_0}, we find that $\Im(p^2 q) = 0$ when
\begin{equation}
  (1+\alpha)(3+4\alpha) x^2 - 8 (1+\alpha) x + 6 = 0\+,
\end{equation}
to leading order in $1/P$.
This equation only has a solution with $0 < x < (1+\alpha)^{-1}$ when $\alpha < -1/4$ but in that case one can verify that $p^2 q$ is real \emph{positive}, not negative, at this point.
Hence, $\arg q + 2 \arg p < \pi$ is satisfied along $\gamma_\mathsf{e}$ in all cases.

\begin{figure}[p]
  \centering
  \subfloat[$\alpha = 3$]{\includegraphics[width=230pt]{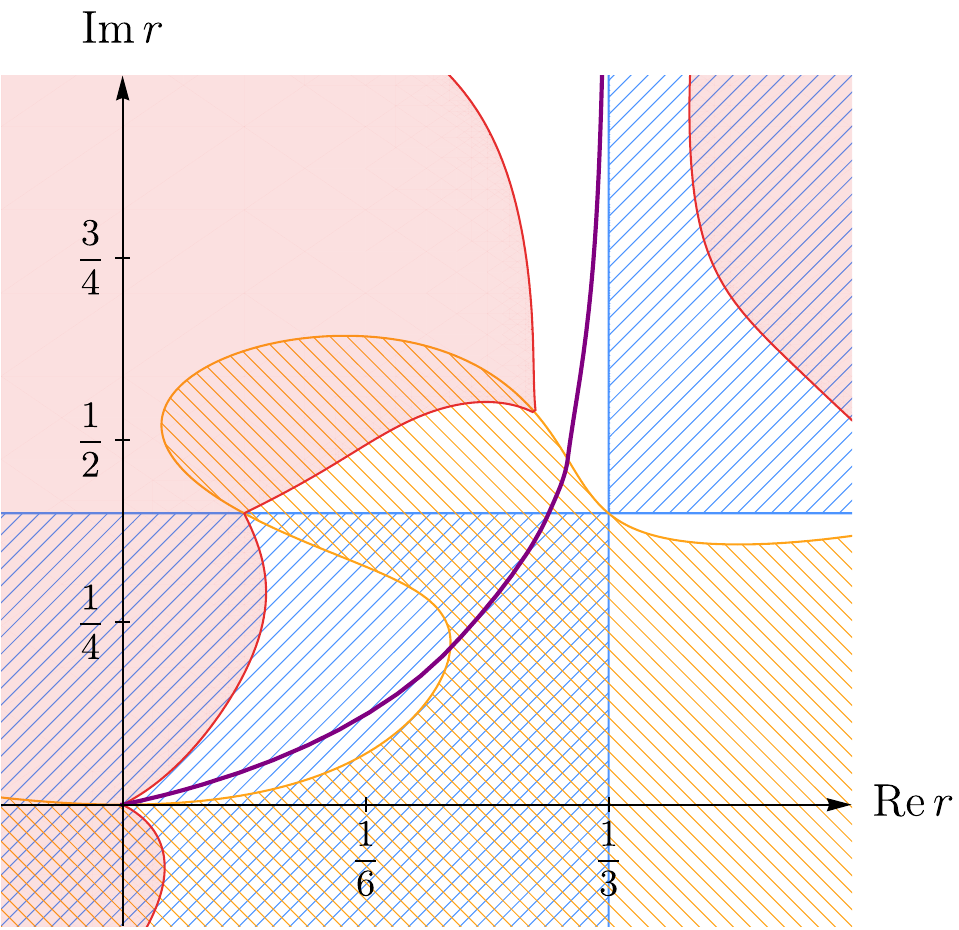}}
  \hfill
  \subfloat[$\alpha = 4.8$]{\includegraphics[width=230pt]{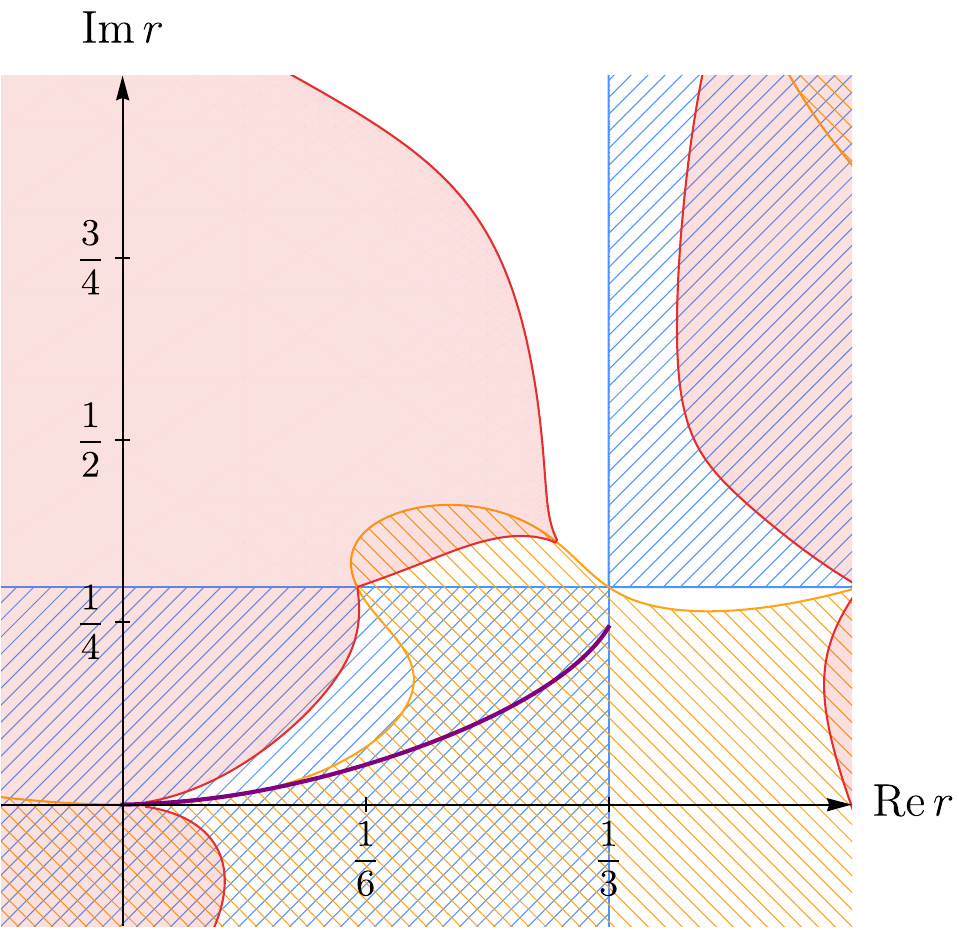}}
  \caption{Illustration of the different behaviors of the extremal curve (purple lines, obtained numerically) depending on whether $\alpha < \alpha_3 \approx 4.530$ or $\alpha > \alpha_3$ for plus-Bolt solutions with $P \to \infty$.
    The colored regions indicate where $\arg p \leq 0$ (blue NE-SW-striped regions), $\arg q \leq 0$ (orange NW-SE-striped regions) and $2\abs{\arg p} + \abs{\arg q} \geq \pi$ (red regions), outside which the naive extremal curve is valid.
    In the $\alpha = 3$ case, the extremal curve escapes to the region where the naive extremal curve is valid close to $\Re r = \Re v = 1/3$ and we conclude that the saddle is allowable whereas, in the $\alpha = 4.8$ case, it does not and the saddle is not allowable.}
  \label{fig:bb9_Bolt_examples}
\end{figure}

The analysis of the extremal curve for the Taub-Bolt-dS saddles is analogous to that of the NUTs although more cumbersome due to lengthier expressions.
The transition for the plus-Bolt at $\alpha_2$, defined in \eqref{eq:bb9_bolt_alpha2}, is similar to the one for the NUT solution at $\alpha = -1/6$.
When $\alpha$ is slightly less than $\alpha_2$, the extremal curve passes below $v$ at a distance from the origin that diverges as $\alpha \to \alpha_2$ for the $P \to \infty$ solutions.
The plus-Bolt transition at $\alpha_3 \approx 4.530$ is of a different type, however.
Here, the question is whether the extremal curve escapes the region close to the origin where $\arg p > 0$, $\arg q > 0$ or $2 \abs{\arg p} + \abs{\arg q} < \pi$ might be violated.
In contrast to the transition at $\alpha_2$, this can be diagnosed at a finite distance from the origin in the $P \to \infty$ solutions.
This transition is illustrated in Fig.~\ref{fig:bb9_Bolt_examples}.

\begin{figure}[p]
  \centering
  \subfloat[NUT]{\includegraphics[width=230pt]{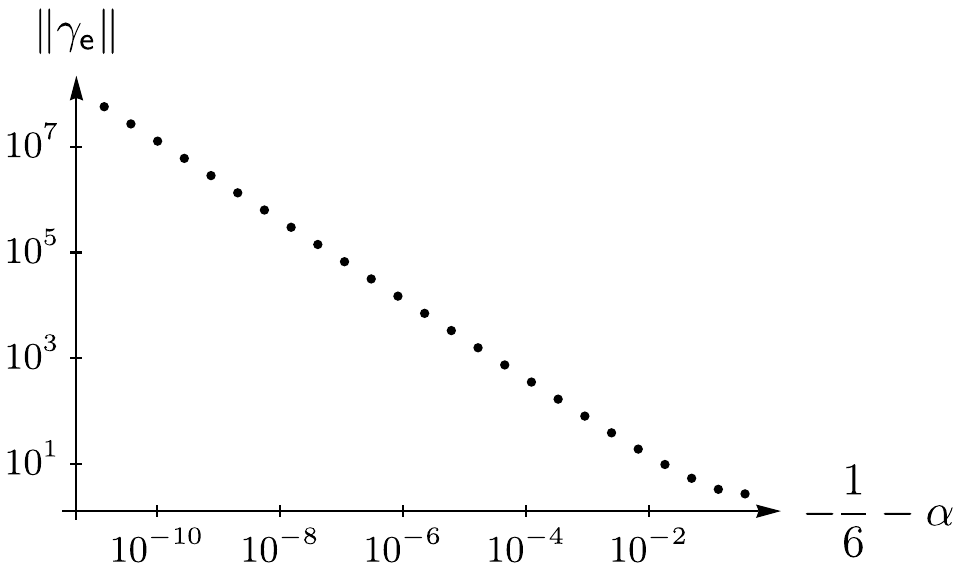}}
  \hfill
  \subfloat[plus-Bolt]{\includegraphics[width=230pt]{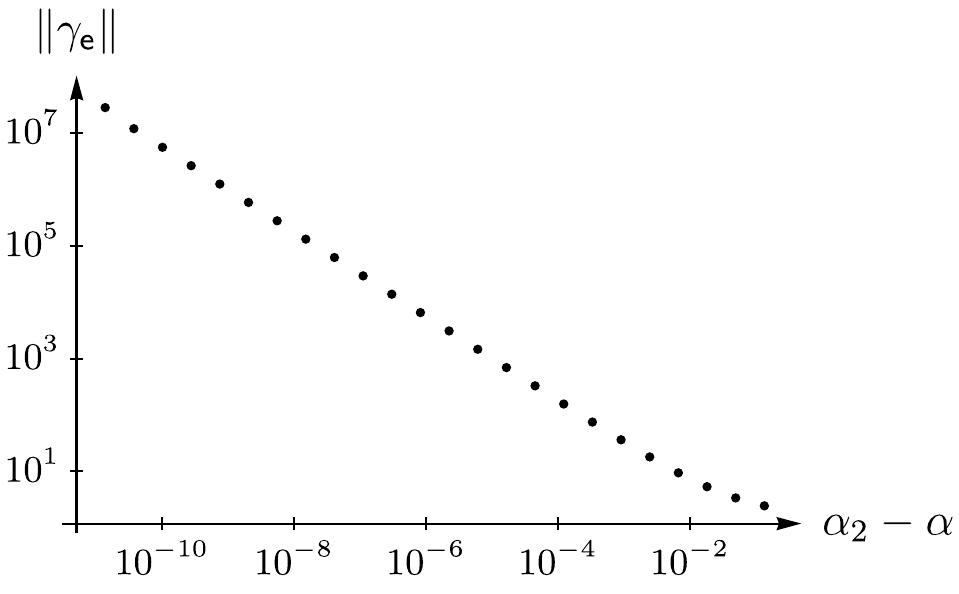}}
  \caption{Numerical evidence that the transitions to allowable saddles happen at $-1/6$ and $\alpha_2$ (defined in \eqref{eq:bb9_bolt_alpha2}) for the NUTs and plus-Bolts, respectively.
    Here, $\|\gamma_\mathsf{e}\|$ is the length of the extremal curve from the origin until $\Re \gamma_\mathsf{e} = \Re v$ for the $P \to \infty$ saddles.
    Since $\|\gamma_\mathsf{e}\| < \infty$ for the points shown, the corresponding saddles are not allowable.
    The diverging power laws $\|\gamma_\mathsf{e}\| \propto (-1/6 - \alpha)^{-3/4}$ and $\|\gamma_\mathsf{e}\| \propto (\alpha_2 - \alpha)^{-3/4}$ fit the two sets of data well close to $-1/6$ and $\alpha_2$, respectively.}
  \label{fig:numerics}
\end{figure}

Except for the precise location of the $\alpha_3$ transition, there are two ranges of $\alpha$ where we resort to numerics.
The first one is for $-1 < \alpha < -1/6$ in the NUT solution and the second for $-0.468 \approx \alpha_1 < \alpha < \alpha_2 \approx -0.216$ in the plus-Bolt solution.
In both cases, the naive extremal curve is only valid up to a point where $q \in \mathbb{R}$ at an $\mathcal{O}(P^0)$ distance from the origin.
Taking a solution far from the upper bound of these ranges, like the one in Fig.~\hyperref[fig:bb9_NUT_examples]{\ref*{fig:bb9_NUT_examples}.a}, one can deduce from the extremal curve that the saddles are not allowable.
To see that the transitions happen precisely at $\alpha = 1/6$ in the NUT case and $\alpha = \alpha_2$ in the plus-Bolt case we approach these values from below while tracking the length $\|\gamma_\mathsf{e}\|$ of the extremal curve from the origin to $\Re \gamma_\mathsf{e} = \Re v$ in the strict $P \to \infty$ limit.
The result shows that $\|\gamma_\mathsf{e}\|$ diverges as a power law at $\alpha = -1/6$ and $\alpha = \alpha_2$ in the two cases, respectively, see Fig.~\ref{fig:numerics}.

\newpage
\bibliographystyle{klebphys2}
\bibliography{refs}

\end{document}